\documentclass[10pt,draftclsnofoot,onecolumn,nofonttune]{IEEEtran}

\usepackage[utf8]{inputenc}

\usepackage{amsmath}
\usepackage{amssymb}
\usepackage{amsfonts}
\usepackage{amsthm}

\usepackage{paralist}
\usepackage{xifthen}

\newlength{\figWidthOne}
\newlength{\figWidthTwo}
\newlength{\figWidthFor}
\newlength{\figWidthThr}

\ifCLASSOPTIONdraftcls
\else
\fi

\usepackage{array}
\usepackage{tabularx}
\usepackage[caption=false,font=footnotesize]{subfig}
\usepackage{url}
\usepackage{xfrac}
\usepackage[algoruled, vlined, linesnumbered]{algorithm2e}

\usepackage{txfonts}
\let\mathbb\varmathbb
\usepackage[mathcal]{euscript}

\usepackage[dvipsnames,svgnames]{xcolor}
\colorlet{MyBlue}{DodgerBlue!50!Black}

\usepackage{lipsum}

\usepackage{hyperref}
\hypersetup{
	draft=false,
	colorlinks=true,
	linktocpage=true,
	pdfstartview=FitH,
	breaklinks=true,
	pdfpagemode=UseNone,
	pageanchor=true,
	pdfpagemode=UseOutlines,
	plainpages=false,
	bookmarksnumbered,
	bookmarksopen=false,
	bookmarksopenlevel=1,
	hypertexnames=true,
	pdfhighlight=/O,
	urlcolor=Maroon,linkcolor=MyBlue,citecolor=DarkGreen, 
	pdftitle={},
	pdfauthor={},
	pdfsubject={},
	pdfkeywords={},
	pdfcreator={pdfLaTeX},
	pdfproducer={LaTeX with hyperref}
}

\usepackage{tikz}
\usetikzlibrary{shapes,arrows,dsp,chains}

\usepackage{pgfplots}
\pgfplotsset{compat=newest}
\pgfplotsset{plot coordinates/math parser=false}

\usepackage{balance}
\usepackage[numbers,sort&compress]{natbib}

\usepackage{placeins}

\theoremstyle{plain}
\newtheorem{theorem}{Theorem}

\newtheorem*{corollary*}{Corollary}

\newtheorem{proposition}[theorem]{Proposition}

\theoremstyle{definition}

\newtheorem*{definition*}{Definition}

\theoremstyle{remark}
\newtheorem{remark}{Remark}
\newtheorem*{remark*}{Remark}

\newcommand{\fC}{\mathbb{C}}

\newcommand{\fR}{\mathbb{R}}

\DeclareMathOperator{\Reg}{Reg}

\newcommand{\inPar}[1]{\left(#1\right)}
\newcommand{\inBra}[1]{\left[#1\right]}
\newcommand{\inCBra}[1]{\left\{#1\right\}}

\newcommand{\bvec}[1]{\mathbf{#1}}
\newcommand{\bmat}[1]{\mathbf{#1}}
\newcommand{\conjT}[1]{#1^{\dag}}

\newcommand{\eye}{\bmat{I}}

\DeclareMathOperator{\modiag}{diag}
\DeclareMathOperator{\motr}{tr}
\newcommand{\diag}[1]{\modiag\inPar{#1}}
\newcommand{\tr}[1]{\motr\inBra{#1}}

\newcommand{\insum}{\sum\nolimits}

\newcommand{\abs}[1]{\left|#1\right|}
\newcommand{\norm}[1]{\left\|#1\right\|}

\newcommand{\smallnorm}[1]{\|#1\|}

\newcommand{\ceil}[1]{\left\lceil #1 \right\rceil}
\newcommand{\floor}[1]{\left\lfloor #1 \right\rfloor}

\newcommand{\der}[3][]{
  \ifthenelse{\isempty{#1}}{
    \frac{\:d{#2}}{\:d{#3}}
  }{
    \frac{\:d^{#1}{#2}}{\:d{#3}^{#1}}}
}

\newcommand{\partDer}[4][]{
  \ifthenelse{\isempty{#1}}{
    \ifthenelse{\isempty{#4}}{
      \frac{\:\partial{#2}}{\:\partial{#3}}
    }{
      \left.\frac{\:\partial{#2}}{\:\partial{#3}}\right|_{#3=#4}
    }
  }{
    \ifthenelse{\isempty{#4}}{
      \frac{\:\partial^{#1}{#2}}{\:\partial{#3}^{#1}}
    }{
      \left.\frac{\:\partial^{#1}{#2}}{\:\partial{#3}^{#1}}\right|_{#3=#4}
    }
  }
}

\newcommand{\expectation}[1]{\mathbb{E}\inBra{#1}}

\newcommand{\ctime}{t}
\newcommand{\ctimeAux}{s}
\newcommand{\dtime}{n}
\newcommand{\dtimeAux}{m}
\newcommand{\timeHorizon}{T}

\newcommand{\connection}{u}
\newcommand{\interfConnection}{v}
\newcommand{\nOfConnections}{U}
\newcommand{\connections}{\mathcal{\nOfConnections}}

\newcommand{\subcarrier}{k}
\newcommand{\nOfSubcarriers}{K}
\newcommand{\subcarriers}{\mathcal{\nOfSubcarriers}}

\newcommand{\tx}[1]{t_{#1}}
\newcommand{\rx}[1]{r_{#1}}

\newcommand{\nOfTx}[1][]{
  \ifthenelse{\isempty{#1}}{
    M
  }{
    M_{#1}
  }
}
\newcommand{\nOfRx}[1][]{
  \ifthenelse{\isempty{#1}}{
    N
  }{
    N_{#1}
  }
}

\newcommand{\tSgn}[2][]{
  \ifthenelse{\isempty{#1}}{
    \bvec{x}_{#2}
  }{
    \bvec{x}^{#1}_{#2}
  }
}

\newcommand{\rSgn}[2][]{
  \ifthenelse{\isempty{#1}}{
    \bvec{y}_{#2}
  }{
    \bvec{y}^{#1}_{#2}
  }
}

\newcommand{\noiseSgn}[2][]{
  \ifthenelse{\isempty{#1}}{
    \bvec{z}_{#2}
  }{
    \bmat{z}^{#1}_{#2}
  }
}

\newcommand{\interfSgn}[2][]{
  \ifthenelse{\isempty{#1}}{
    \bvec{w}_{#2}
  }{
    \bmat{w}^{#1}_{#2}
  }
}

\newcommand{\channel}[2][]{
  \ifthenelse{\isempty{#1}}{
    \bmat{H}_{#2}
  }{
    \bmat{H}^{#1}_{#2}
  }
}

\newcommand{\interfPower}[2][]{
  \ifthenelse{\isempty{#1}}{
    \bmat{W}_{#2}
  }{
    \bmat{W}^{#1}_{#2}
  }
}

\newcommand{\effChannel}[2][]{
  \ifthenelse{\isempty{#1}}{
    \ifthenelse{\isempty{#2}}{
      \widetilde{\bmat{H}}
    }{
      \widetilde{\bmat{H}}(#2)
    }
  }{
    \ifthenelse{\isempty{#2}}{
      \widetilde{\bmat{H}}_{#1}
    }{
      \widetilde{\bmat{H}}_{#1}(#2)
    }
  }
}

\newcommand{\effChannelCT}[2][]{
  \ifthenelse{\isempty{#1}}{
    \ifthenelse{\isempty{#2}}{
      \conjT{\widetilde{\bmat{H}}}
    }{
      \conjT{\widetilde{\bmat{H}}}(#2)
    }
  }{
    \ifthenelse{\isempty{#2}}{
      \conjT{\widetilde{\bmat{H}}}_{#1}
    }{
      \conjT{\widetilde{\bmat{H}}}_{#1}(#2)
    }
  }
}

\newcommand{\power}[2][]{
  \ifthenelse{\isempty{#1}}{
    \ifthenelse{\isempty{#2}}{
      \bmat{Q}
    }{
      \bmat{Q}(#2)
    }
  }{
    \ifthenelse{\isempty{#2}}{
      \bmat{Q}_{#1}
    }{
      \bmat{Q}_{#1}(#2)
    }
  }
}

\newcommand{\powerC}[2][]{
  \ifthenelse{\isempty{#1}}{
    \ifthenelse{\isempty{#2}}{
      \bmat{Q}^{c}
    }{
      \bmat{Q}^{c}(#2)
    }
  }{
    \ifthenelse{\isempty{#2}}{
      \bmat{Q}^{c}_{#1}
    }{
      \bmat{Q}^{c}_{#1}(#2)
    }
  }
}

\newcommand{\powerDens}[2][]{
  \ifthenelse{\isempty{#1}}{
    \ifthenelse{\isempty{#2}}{
      \bmat{D}
    }{
      \bmat{D}\inPar{#2}
    }
  }{
    \ifthenelse{\isempty{#2}}{
      \bmat{D}_{#1}
    }{
      \bmat{D}_{#1}\inPar{#2}
    }
  }
}

\newcommand{\pmax}[1][]{
  \ifthenelse{\isempty{#1}}
  {
    P
  }{
    P_{#1}
  }
}

\newcommand{\fxPower}[1][]{
  \ifthenelse{\isempty{#1}}{
    \bmat{Q}^{\ast}
  }{
    \bmat{Q}^{\ast}_{#1}}
}

\newcommand{\fxPowerDens}[1][]{
  \ifthenelse{\isempty{#1}}{
    \bmat{D}^{\ast}
  }{
    \bmat{D}^{\ast}_{#1}
  }
}

\newcommand{\loss}[2]{
  \ell(#1;#2)
}

\newcommand{\lossPM}[2][]{
  \ifthenelse{\isempty{#1}}{
    \ifthenelse{\isempty{#2}}{
      \ell
    }{
      \ell(#2)
    }
  }{
    \ifthenelse{\isempty{#2}}{
      \ell(#1)
    }{
      \ell(#1;#2)
    }
  }
}

\newcommand{\rate}[2]{
  \ifthenelse{\isempty{#2}}{
    R(#1)
  }{
    R(#1;#2)
  }
}

\newcommand{\tRate}[2][]{
  \ifthenelse{\isempty{#1}}{
    \ifthenelse{\isempty{#2}}{
      R^{\ast}
    }{
      R^{\ast}(#2)
    }
  }{
    \ifthenelse{\isempty{#2}}{
      R^{\ast}_{#1}
    }{
      R^{\ast}_{#1}(#2)
    }
  }
}

\newcommand{\etaSeq}[1]{\eta(#1)}
\newcommand{\etaSqrt}[1]{\eta{#1}^{-1/2}}

\newcommand{\etaSeqC}[1]{\eta^{c}(#1)}

\newcommand{\gradient}[2][]{
  \ifthenelse{\isempty{#1}}{
    \ifthenelse{\isempty{#2}}{
      \bmat{V}
    }{
      \bmat{V}(#2)
    }
  }{
    \ifthenelse{\isempty{#2}}{
      \bmat{V}_{#1}
    }{
      \bmat{V}_{#1}(#2)
    }
  }
}

\newcommand{\gradientC}[2][]{
  \ifthenelse{\isempty{#1}}{
    \ifthenelse{\isempty{#2}}{
      \bmat{V}^{c}
    }{
      \bmat{V}^{c}(#2)
    }
  }{
    \ifthenelse{\isempty{#2}}{
      \bmat{V}_{#1}^{c}
    }{
      \bmat{V}_{#1}^{c}(#2)
    }
  }
}

\newcommand{\gradientN}[2][]{
  \ifthenelse{\isempty{#1}}{
    \ifthenelse{\isempty{#2}}{
      \bmat{\hat{V}}
    }{
      \bmat{\hat{V}}(#2)
    }
  }{
    \ifthenelse{\isempty{#2}}{
      \bmat{\hat{V}}_{#1}
    }{
      \bmat{\hat{V}}_{#1}(#2)
    }
  }
}

\newcommand{\gradientB}[1][]{
  \ifthenelse{\isempty{#1}}{
    V
  }{
    V_{#1}
  }
}

\newcommand{\noiseGradient}[2][]{
  \ifthenelse{\isempty{#1}}{
    \ifthenelse{\isempty{#2}}{
      \bmat{Z}
    }{
      \bmat{Z}(#2)
    }
  }{
    \ifthenelse{\isempty{#2}}{
      \bmat{Z}_{#1}
    }{
      \bmat{Z}_{#1}(#2)
    }
  }
}

\newcommand{\noiseGradientB}[1][]{
  \ifthenelse{\isempty{#1}}{
    \Sigma
  }{
    \Sigma_{#1}
  }
}

\newcommand{\cumGradient}[2][]{
  \ifthenelse{\isempty{#1}}{
    \ifthenelse{\isempty{#2}}{
      \bmat{Y}
    }{
      \bmat{Y}(#2)
    }
  }{
    \ifthenelse{\isempty{#2}}{
      \bmat{Y}_{#1}
    }{
      \bmat{Y}_{#1}(#2)
    }
  }
}

\newcommand{\cumGradientC}[2][]{
  \ifthenelse{\isempty{#1}}{
    \ifthenelse{\isempty{#2}}{
      \bmat{Y}^{c}
    }{
      \bmat{Y}^{c}(#2)
    }
  }{
    \ifthenelse{\isempty{#2}}{
      \bmat{Y}_{#1}^{c}
    }{
      \bmat{Y}_{#1}^{c}(#2)
    }
  }
}

\newcommand{\regret}[2]{
  \Reg(#1;#2)
}

\newcommand{\bA}{\mathbf{A}}

\newcommand{\bH}{\mathbf{H}}

\newcommand{\bQ}{\mathbf{Q}}

\newcommand{\bU}{\mathbf{U}}
\newcommand{\bV}{\mathbf{V}}

\newcommand{\bX}{\mathbf{X}}
\newcommand{\bY}{\mathbf{Y}}

\DeclareMathOperator{\prob}{\mathbb{P}}
\DeclareMathOperator{\ex}{\mathbb{E}}
\DeclareMathOperator{\bigoh}{\mathcal O}

\newcommand{\eps}{\varepsilon}
\newcommand{\from}{\colon}
\newcommand{\state}{\mathcal{X}}
\newcommand{\eq}{\bX^{\ast}}
\newcommand{\temp}{\eta}
\newcommand{\dd}{\:d}

\newcommand{\mg}{\succ}
\newcommand{\mgeq}{\succcurlyeq}

\usepackage{acronym}

\newacro{FM}{Fo\-schi\-ni\textendash Mil\-ja\-nic}
\newacro{5G}{5th generation}
\newacro{CCI}{co-channel interference}
\newacro{MUI}{multi-user interference-plus-noise}
\newacro{SINR}{signal-to-interference-and-noise ratio}
\newacro{PC}{power control}
\newacro{OGD}{online gradient descent}
\newacro{OMD}{online mirror descent}
\newacro{MIMO}{multiple-input and multiple-output}
\newacro{SISO}{single-input and single-output}
\newacro{OFDM}{orthogonal frequency-division multiplexing}
\newacro{OFDMA}{orthogonal frequency-division multiple access}
\newacro{XL}{exponential learning}
\newacro{MXL}{matrix exponential learning}
\newacro{CSI}{channel state information}
\newacro{CSIT}{channel state information at the transmitter}
\newacro{QoS}{quality of service}
\newacro{QoE}{quality of experience}
\newacro{TDD}{time-division duplexing}
\newacro{DL}{downlink}
\newacro{UL}{uplink}
\newacro{EPA}{extended pedestrian A}
\newacro{EVA}{extended vehicular A}
\newacro{ETU}{extended typical urban}
\newacro{CR}{cognitive radio}
\newacro{FSI}{Freund and Schapire informed}
\newacro{MAC}{multiple access channel}
\newacro{iid}[i.i.d.]{independent and identically distributed}
\newacro{SUD}{single user decoding}
\newacro{MSE}{mean square error}
\newacro{BS}{base station}

\begin{document}

\title{Adaptive Power Allocation and Control in Time-Varying Multi-Carrier MIMO Networks}

\author{%
Ioannis~Stiakogiannakis,
\IEEEmembership{Member,~IEEE,}
Panayotis~Mertikopoulos,
\IEEEmembership{Member,~IEEE,}
and
Corinne~Touati%
\thanks{%
This research was supported in part
by the French National Research Agency project NETLEARN (ANR\textendash 13\textendash INFR\textendash 004)
and
by the European Commission in the framework of the FP7 Network of Excellence in Wireless COMmunications NEWCOM\# (contract no. 318306).
Part of this work was presented in Allerton 2014.}
\thanks{%
Ioannis Stiakogiannakis is with the Mathematical and Algorithmic Sciences Lab, France Research Center, Huawei Technologies Co. Ltd; this work was completed while he was with Inria, France.
Panayotis Mertikopoulos is with the French National Center for Scientific Research (CNRS) and the Univ. Grenoble Alpes, LIG, F-38000 Grenoble, France;
Corinne Touati is with Inria, France, and the Laboratoire d'Informatique de Grenoble, LIG, F-38000 Grenoble, France.}
\thanks{Email: ioannis.steiakogiannakis@huawei.com, panayotis.mertikopoulos@imag.fr, corinne.touati@inria.fr}
\thanks{Manuscript received \today.}} 

\maketitle

\begin{abstract}
In this paper, we examine the fundamental trade-off between radiated power and achieved throughput in wireless multi-carrier, \ac{MIMO} systems that vary with time in an unpredictable fashion (e.g. due to changes in the wireless medium or the users' \acs{QoS} requirements).
Contrary to the static/stationary channel regime, there is no optimal power allocation profile to target (either static or in the mean), so the system's users must adapt to changes in the environment ``on the fly'', without being able to predict the system's evolution ahead of time.
In this dynamic context, we formulate the users' power/throughput trade-off as an online optimization problem, and we provide a matrix exponential learning algorithm that leads to \emph{no regret} \textendash\ i.e. the proposed transmit policy is asymptotically optimal in hindsight, irrespective of how the system evolves over time.
Furthermore, we also examine the robustness of the proposed algorithm under imperfect \ac{CSI} and we show that it retains its regret minimization properties under very mild conditions on the measurement noise statistics.
As a result, users are able to track the evolution of their individually optimum transmit profiles remarkably well, even under rapidly changing network conditions and high uncertainty.
Our theoretical analysis is validated by extensive numerical simulations corresponding to a realistic network deployment, and providing further insights in the practical implementation aspects of the proposed algorithm.
\end{abstract}


\IEEEpeerreviewmaketitle


\section{Introduction}
\label{sec:Introduction}


The wildfire spread of Internet-enabled mobile devices is putting existing wireless systems under enormous strain and is one of the driving forces behind the transition to next-generation mobile networks \cite{ABC+14}.
In this context, the efficient control and allocation of radiated power comprises an indispensable aspect of wireless system design:
for many applications (such as e-mail and voice calls), radiated power must be reduced to the bare minimum in order to preserve battery life;
by contrast, for rate-hungry applications (such as multimedia streaming and video calling), it is crucial to optimize the allocation of the users' limited power across the network's degrees of freedom so as to maximize their throughput.
In this way, wireless users are facing an important trade-off between radiated power and achieved throughput which must often be resolved in an adaptive and distributed manner, with minimal coordination between users.

In its most basic form, \ac{PC} allows wireless links to achieve a target throughput while minimizing radiated power and the induced \ac{CCI}.
Accordingly, \acl{PC} has had a pivotal impact on wireless system design and operation ever since the early development stages of legacy wireless networks:
starting with the pioneering work of Zander \cite{zander_performance_1992}, Grandhi et al. \cite{grandhi_distributed_1994}, Foschini and Miljanic \cite{foschini_simple_1993} and Yates \cite{yates_framework_1995}, the design of efficient \acl{PC} algorithms has given rise to a vast and extremely active corpus of literature \textendash\ see e.g. \cite{CHLT07} for a survey.
Thus, in view of recent advances in \ac{MIMO} technologies and the prolific deployment of \ac{OFDMA} schemes, the envisioned transition to \ac{5G} mobile systems calls for \acl{PC} algorithms tailored to networks with several degrees of freedom (spectral as well as spatial).

In this setting, most of the relevant literature has focused on maximizing the users' achievable transmission rate subject to their individual power constraints:
\cite{shen_optimal_2003,YRBC04,kobayashi_iterative_2006} treat rate maximization as a constrained nonlinear optimization problem whereas \cite{SPB09-sp,YSSP13,BLD09} focus on multiple user interactions using game-theoretic methods;
in a similar vein, \cite{oh_optimum_2004, yu_spc10-2:_2006, wunder_optimal_2007} studied the power minimization problem subject to the users' rate requirements in multi-carrier \acp{MAC}, while \cite{MSW14} provided a two-layer framework for power minimization in \ac{MIMO}\textendash\ac{OFDMA} systems.
However, while the benefits of \acl{PC} algorithms are relatively easy to assess in static networks, it is much harder to analyze their behavior in wireless systems that vary with time (e.g. due to user mobility, fading, temporal variations in the wireless medium, etc.).
In the ergodic regime (where the users' channels follow a stationary ergodic process), \cite{BCP00,holliday_distributed_2003} provided \acl{PC} algorithms that minimize the users' transmit power while achieving a minimum ergodic rate requirement.
More recently, the authors of \cite{MBML12} studied the problem of ergodic rate maximization in fast-fading multi-carrier systems and they provided an efficient power allocation algorithm that allows users to attain the system's (ergodic) sum-capacity.
However, when the wireless medium does not evolve according to an \ac{iid} sequence of random variables,
the efficient allocation and control of radiated power remains a very open issue.

In this paper, we drop all stationarity/\ac{iid} assumptions and we focus squarely on wireless systems that evolve \emph{arbitrarily} over time in terms of both channel conditions and user \ac{QoS} requirements.
In this framework, standard approaches based on linear programming (for static channels) and/or stochastic optimization (for the ergodic regime) are no longer relevant because there is no underlying optimization problem to solve \textendash\ either static or in the mean.
Instead, we treat power control as a \emph{dynamically evolving optimization problem} and we employ techniques and ideas from online learning and optimization \cite{SS11} to quantify how well the system's users can adapt to changes in the wireless medium.

The most widely used performance criterion in this setting is that of \emph{regret minimization}, a seminal concept which was first introduced by Hannan \cite{Han57} and which has since given rise to a vigorous literature at the interface of machine learning, optimization, statistics, and game theory \textendash\ for a comprehensive survey, see e.g. \cite{CBL06,SS11}.
Specifically, in the language of game theory, the notion of regret compares a user's cumulative payoff over a given time horizon to the cumulative payoff that he would have obtained by employing the \emph{a posteriori} best possible action over the time horizon in question.
Accordingly, in the context of power allocation and control, regret minimization corresponds to dynamic transmit policies that are asymptotically optimal in hindsight, \emph{irrespective} of how the user's environment and/or requirements evolve over time.

Regret minimization was recently used in \cite{dams_convergence_2012} to study the transient phase of the \ac{FM} \acl{PC} algorithm in static environments and to propose alternative convergent \acl{PC} schemes based on the notion of \emph{swap regret} \cite{blum_external_2007}.
In \cite{latifa_no-regret_2012}, the authors considered a potential game formulation for the joint \acl{PC} and channel allocation problem in \ac{CR} networks and they employed a regret minimizing algorithm \cite{auer_gambling_1995} to reach a Nash equilibrium state.
The same problem was also examined in the context of infrastructureless wireless networks by the authors of \cite{maghsudi_joint_2014} who formulated the problem as a potential game and provided a \acl{PC} algorithm based on internal regret minimization that converges to the game's unique correlated \textendash\ and, hence, Nash \textendash\ equilibrium.
Finally, in a very recent paper, the authors of \cite{MB14} employed online optimization methodologies to derive a dynamic transmit policy for online rate maximization in \acl{CR} networks, but without attempting to control the users' radiated power level.

\subsection*{Summary of results and paper outline}

In this paper, we focus on multi-user \ac{MIMO}\textendash\ac{OFDMA} systems that evolve arbitrarily over time (for instance, due to fading, intermittent user activity, changing \ac{QoS} requirements, etc.), and we seek to provide an efficient \acl{PC} and allocation scheme that allows users to balance their radiated power against their achieved throughput ``on the fly'', based only on locally available (and possibly imperfect) \ac{CSI}.
In particular, we formulate the wireless users' power minimization/throughput maximization trade-off as an online optimization problem and we derive a \emph{no-regret} power control policy based on the method of \ac{MXL} \cite{TRW05,MBM12,KSST12}.
The proposed \ac{MXL} algorithm is provably asymptotically optimal against the system's evolution in hindsight;
furthermore, it also enjoys the following desirable properties:
\begin{itemize}
\item
\emph{Distributedness:}
users update their own power profiles based only on local information.
\item
\emph{Asynchronicity:}
there is no need for a global update timer to synchronize user updates.
\item
\emph{Robustness:}
the algorithm retains its properties even under imperfect \ac{CSI}.
\item
\emph{Statelessness:}
transmitters do not need to know the network's state and/or topology.
\end{itemize}

This work builds on (and significantly extends) our recent results on the regret minimization properties of the original \acl{FM} dynamics in \ac{SISO}, single-carrier systems that evolve continuously over time \cite{stiakogiannakis_no_2014}.
Compared to \cite{stiakogiannakis_no_2014}, the current paper represents an extension to multi-carrier systems with several antennas (at both the transmitter and the receiver) and with imperfect feedback and \ac{CSIT}.


After presenting our wireless system model in Section \ref{sec:SystemModelAndProblemFormulation}, the proposed algorithm for adaptive \acl{PC} in \ac{MIMO}\textendash\ac{OFDMA} systems is derived in Section \ref{sec:PowerControlForMIMOOFDM}.
Our main result therein is that the proposed algorithm leads to no regret;
in addition, we examine the algorithm's behavior in the presence of imperfect \ac{CSI} and we show that the algorithm retains its regret minimization properties almost surely, irrespective of the measurement noise level.
Our theoretical analysis is supplemented by extensive numerical simulations in Section \ref{sec:NumericalResults} where we illustrate the power and throughput gains of the proposed \acl{PC} algorithm under realistic network conditions.

\section{System Model and Problem Formulation}
\label{sec:SystemModelAndProblemFormulation}

Consider a set $\connections = \inCBra{1,\ldots,\nOfConnections}$ of wireless point-to-point connections formed over a set of orthogonal subcarriers $\subcarriers = \inCBra{1,\ldots,\nOfSubcarriers}$;
assume further that each connection $\connection\in\connections$ comprises a transmit-receive pair $(\tx{\connection},\rx{\connection})$ with $\nOfTx[\connection]$ antennas at the transmitter and $\nOfRx[\connection]$ antennas at the receiver.
Thus, if $\tSgn[\connection]{\subcarrier}\in\fC^{\nOfTx[\connection]}$ and $\rSgn[\connection]{\subcarrier}\in\fC^{\nOfRx[\connection]}$ denote respectively the signals transmitted and received over connection $\connection$ on subcarrier $\subcarrier$, we obtain the familiar signal model:
\begin{equation}
  \rSgn[\connection]{\subcarrier} = \channel[\connection\connection]{\subcarrier}\tSgn[\connection]{\subcarrier}+\insum_{\interfConnection\neq\connection}{\channel[\interfConnection\connection]{\subcarrier}\tSgn[\interfConnection]{\subcarrier}}+\noiseSgn[\connection]{\subcarrier}
  \label{eq:receivedSignal}
\end{equation}
where
$\noiseSgn[\connection]{\subcarrier}\in\fC^{\nOfRx[\connection]}$ denotes the ambient noise over subcarrier $\subcarrier$ (including thermal, atmospheric and other peripheral interference effects)
and $\channel[\interfConnection\connection]{\subcarrier}\in\fC^{\nOfRx[\connection]\times\nOfTx[\interfConnection]}$ is the transfer matrix between $\tx{\interfConnection}$ and $\rx{\connection}$.

\begin{figure}[t]
\centering

\tikzstyle{source}  = [draw, fill=DodgerBlue!30, rectangle, minimum height=1cm, minimum width=1cm]
\tikzstyle{msource} = [draw, fill=DodgerBlue!30, rectangle, minimum height=1cm, minimum width=1cm, scale=0.75]
\tikzstyle{dest}    = [draw, fill=DodgerBlue!30, circle,    minimum height=1cm, minimum width=1cm]
\tikzstyle{mdest}   = [draw, fill=DodgerBlue!30, circle,    minimum height=1cm, minimum width=1cm, scale=0.75]
\tikzstyle{nonode}  = [draw, fill=white, circle, minimum size=\dspoperatordiameter]

\begin{tikzpicture}[auto, node distance=2cm,>=latex']
  
  \node [cloud, draw, cloud puffs=20, cloud puff arc=60, minimum width=7cm, minimum height=3cm, inner ysep=1em] (cloud) at (1.5,-0.5) {};
  
  \node [nonode, below=-0.215cm of cloud] (noname) {};
  \node [dspadder, below=-0.215cm of cloud] (sum1) {};
  \node [msource, above left=1.6cm and 1.25cm of sum1] (source1) {$\tx{1}$};
  \node [mdest,   above right=1.705cm and 1.25cm of sum1] (destination1) {$\rx{1}$};
  \node [msource, below of=source1,      node distance=1.75cm] (source2) {$\tx{\nOfConnections}$};
  \node [mdest,   below of=destination1, node distance=1.75cm] (destination2) {$\rx{\nOfConnections}$};

  \node [above=0.01mm of source2] (sourcedots) {$\vdots$};
  \node [above=0.01mm of destination2] (destinationdots) {$\vdots$};

  \draw [draw,->] (source1) -- (destination1);
  \draw [draw,->] (source2) -- (destination2);
  \draw [draw,->] (source1) -- (sum1);
  \draw [draw,->] (source2) -- (sum1);

  \node [dspadder, below of=sum1] (sum) {};
  \node [source, node distance=2.5cm, left of=sum] (source) {$\tx{\connection}$};
  \node [dest, node distance=2.5cm, right of=sum] (destination) {$\rx{\connection}$};
  \node [below=2.5cm of sum1] (noise) {$\noiseSgn[\connection]{\subcarrier}$};
  
  \draw [->] (source) -- node {$\channel[\connection\connection]{\subcarrier}\tSgn[\connection]{\subcarrier}$} (sum);
  \draw [->] (sum) -- node {$\rSgn[\connection]{\subcarrier}$} (destination);
  \draw [->] (sum1) -- node [near start] {$\interfSgn[\connection]{\subcarrier} = \sum\nolimits_{\interfConnection\neq\connection}{\channel[\interfConnection\connection]{\subcarrier}\tSgn[\interfConnection]{\subcarrier}}$} (sum);
  \draw [->] (noise) -- (sum);

\end{tikzpicture}
\caption{Example of a network with several active connections where we focus on a particular connection $\connection$ between transmitter $\tx{\connection}$ and receiver $\rx{\connection}$.
The other active connections $\interfConnection\in\connections\setminus\{\connection\}$ cause co-channel interference to the focal connection $\connection\in\connections$ which, together with the ambient subcarrier noise, is treated as additive colored noise.}
\label{fig:NetworkDiagram}
\end{figure}
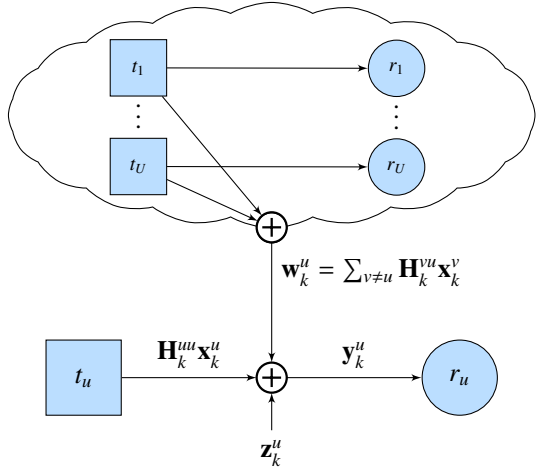

Unavoidably, the received signal $\rSgn[\connection]{\subcarrier}$ is affected by the ambient noise and interference due to the transmissions of other connections on the same subcarrier, so we will write
\begin{equation}
\interfSgn[\connection]{\subcarrier}
	= \insum_{\interfConnection\neq\connection}{\channel[\interfConnection\connection]{\subcarrier}\tSgn[\interfConnection]{\subcarrier}}+\noiseSgn[\connection]{\subcarrier}
\end{equation}
for the \ac{MUI} at the receiver $\rx{\connection}$ of connection $\connection$ (for a schematic representation, see Fig.~\ref{fig:NetworkDiagram});
in this way, \eqref{eq:receivedSignal} attains the simpler form
\begin{equation}
\label{eq:receivedSignalUnified}
\rSgn[\connection]{\subcarrier}
	= \channel[\connection\connection]{\subcarrier}\tSgn[\connection]{\subcarrier}+\interfSgn[\connection]{\subcarrier}.
\end{equation}
In particular, in what follows, we will focus on a specific connection $\connection\in\connections$, so, for clarity, we will drop the index $\connection$ altogether and we will write \eqref{eq:receivedSignalUnified} even more compactly as:
\begin{equation}
\rSgn{\subcarrier}
	= \channel{\subcarrier}\tSgn{\subcarrier}+\interfSgn{\subcarrier}.
\label{eq:receivedSignalUnifiedNoIndex}
\end{equation}

In this context, assuming Gaussian input and noise and \ac{SUD} at the receiver (i.e. the multi-user interference by all other users is treated as additive noise), the transmission rate of the focal connection will be
\cite{telatar_capacity_1999,bolcskei_capacity_2002}:
\begin{equation}
\label{eq:Rate}
\rate{\power{}}{}
	= \insum_{\subcarrier\in\subcarriers}
	\inBra{\log\det\inPar{\interfPower{\subcarrier} + \channel{\subcarrier}\power[\subcarrier]{}\conjT{\channel{\subcarrier}}}
	-\log\det\interfPower{\subcarrier}},
\end{equation}
where $\bH^{\dag}$ denotes the Hermitian conjugate of $\bH$ and:
\begin{itemize}
\item
$\power[\subcarrier]{} = \ex\big[\tSgn{\subcarrier}\conjT{\tSgn{\subcarrier}}\big]$ is the $\nOfTx\times \nOfTx$ covariance matrix of the transmitted signal over subcarrier $\subcarrier$.
\item
$\power{} = \diag{\power[1]{},\ldots,\power[\nOfSubcarriers]{}}$ denotes the power profile of the focal transmitter over all subcarriers.
\item $\interfPower{\subcarrier} = \ex\big[\interfSgn{\subcarrier}\conjT{\interfSgn{\subcarrier}}\big]$ is the $\nOfRx\times\nOfRx$ \ac{MUI} covariance matrix over subcarrier $\subcarrier$.
%
  %
  %
\end{itemize}
In view of the above, let
\begin{equation}
\effChannel[\subcarrier]{}
	= \interfPower{\subcarrier}^{-1/2}\channel{\subcarrier}
\end{equation}
denote the user's \emph{effective channel matrix} over subcarrier $\subcarrier$.
Then, Eq.~\eqref{eq:Rate} can be written as:
\begin{equation}
  \rate{\power{}}{} = \insum_{\subcarrier\in\subcarriers}\log\det\inPar{\eye+\effChannel[\subcarrier]{}\power[\subcarrier]{}\effChannelCT[\subcarrier]{}}
  \label{eq:RateSimplified}
\end{equation}
or, even more concisely:
\begin{equation}
  \rate{\power{}}{} = \log\det\inPar{\eye+\effChannel{}\power{}\effChannelCT{}},
  \label{eq:RateMoreSimplified}
\end{equation}
where the block-diagonal matrix $\effChannel{} = \diag{\effChannel[1]{},\ldots,\effChannel[\nOfSubcarriers]{}}$ collects the user's effective channel matrices over all subcarriers $\subcarrier\in\subcarriers$.

As we mentioned in the introduction, we focus on wireless users who seek to minimize their radiated power on the one hand while maximizing their transmission rate on the other.
Thus, to account for this trade-off between transmit power and achieved throughput, we will consider the general \acl{PC} objective:
\begin{equation}
\lossPM{\power{}}
	= \tr{\power{}}
	- \phi\inPar{\rate{\power{}}{}}
  \label{eq:Loss}
\end{equation}
where $\phi\from\fR_{+}\to\fR$ is a nondecreasing function of the user's achievable transmission rate $\rate{\power{}}{}$.
By this token, $\lossPM{\power{}}$ can be interpreted as a ``\emph{loss function}'' (or negative utility):
higher values of $\lossPM{\power{}}$ indicate that the user is transmitting at very high power, at very low rate, or both, so he is incurring a ``loss''.
Accordingly, we will only assume that $\phi$ is Lipschitz continuous and concave:
the former assumption is a mild technical requirement which we make for simplicity, while the latter reflects the effects of ``diminishing returns'' on ever higher data rates (a rate increase from $1\,\mathrm{bps}$ to $2\,\mathrm{bps}$ is more impactful than an increase from $1,001\,\mathrm{bps}$ to $1,002\,\mathrm{bps}$).

\begin{remark}
Utility-based formulations have a long history in the \acl{PC} literature \textendash\ see e.g. the recent papers \cite{wang_optimal_2014,maghsudi_joint_2014} for a related approach and \cite{saraydar_pricing_2001,MGPS07} for a similar formulation in terms of energy efficiency.
Other possible approaches could involve achieving the Pareto frontier of the dual-objective trade-off between power minimization and throughput maximization;
we focus on the specific model \eqref{eq:Loss} on account of the model's flexibility, generality and overall simplicity.
\end{remark}

\begin{remark}
\label{rem:rate-req}
An important special case of the objective \eqref{eq:Loss} concerns the scenario where the focal user seeks to minimize his transmit power $\tr{\power{}}$ subject to achieving a target transmission rate $\tRate{}$.
This classical formulation of \acl{PC} can be recovered by considering a rate-adjustment function $\phi$ of the form $\phi(R) = f\big(\tRate{}- R\big)$ with $f(r) = 0$ if $r\leq0$ and $f(r)<0$ otherwise \textendash\ for instance, a standard choice would be to take $\phi(R) = -\lambda\cdot \big[\tRate{} - R\big]_{+}$ for some $\lambda>0$.
In this way, when the target transmission rate is achieved (i.e. $\rate{\power{}}{} \geq \tRate{}$), the only term in the user's loss function \eqref{eq:Loss} is the user's total transmit power $\tr{\power{}}$;
otherwise, if the target transmission rate is not met, the user incurs an additional loss of at least $\phi'(0^{-})\cdot \big(\tRate{} - \rate{\power{}}{}\big)$.%
\footnote{Recall here that $\phi$ is assumed concave, so the user's loss grows at least linearly with the rate deficit $\tRate{} - \rate{\power{}}{}$.}
In this way, the (positive) factor $\lambda = \phi'(0^{-})$ represents the tolerance of the connection with respect to transmission rate deficits:
smaller values of $\lambda$ correspond to softer rate requirements, while, in the large $\lambda$ limit, the loss function \eqref{eq:Loss} stiffens to a hard constraint where no violations are tolerated.
\end{remark}

In the above formulation, all sources of noise and \acl{CCI} by other users are collected in the effective channel matrix $\effChannel[\subcarrier]{}$ of the focal connection;
in this way, $\effChannel[\subcarrier]{}$ collects all variables that are not under the direct control of the focal transmitter/receiver pair.
As such, given that we make no assumptions on the behavior of the other connections in the network (or the evolution of the wireless medium itself), the matrix $\effChannel{}$ may vary arbitrarily over time;
our only assumptions will be as follows:
\begin{enumerate}
[({A}1)]
\item $\effChannel{}$ remains bounded for all time (e.g. due to RF circuit losses, antenna directivity, minimum distance between transmitter and receiver, etc.).
\item The variability of $\effChannel{}$ is such that standard results from information theory remain valid \cite{telatar_capacity_1999}.
\end{enumerate}

In this time-varying context, the throughput expression \eqref{eq:Rate} becomes:
\begin{equation}
  \rate{\power{}}{\ctime} = \log\det\inBra{\eye + \effChannel{\ctime}\,\power{}\,\effChannelCT{\ctime}},
  \label{eq:RateSimplifiedTV}
\end{equation}
where $\effChannel{\ctime}$ denotes the user's effective channel matrix at time $\ctime$.
With this in mind, the user's loss function at time $\ctime$ will be
\begin{equation}
\loss{\power{}}{\ctime}
	= \tr{\power{}}
	- \phi(\rate{\power{}}{\ctime};\ctime).
\end{equation}
We thus obtain the following online \acl{PC} problem for \ac{MIMO}\textendash\ac{OFDMA} systems:
\begin{equation}
\begin{aligned}
\text{minimize}
	&\quad \loss{\power{}}{\ctime},
	\\
	\text{subject to}
	&\quad
	\power{} \in \state,
	\\
\end{aligned}
\label{eq:OPC}
\tag{OPC}
\end{equation}
where
\begin{equation}
\state
	= \{\power{}: \power{} \succcurlyeq 0,\; \tr{\power{}} \leq \pmax\}
\end{equation}
is the problem's state space and $\pmax>0$ denotes the user's maximum transmit power.
More precisely, given that the user has no control over the effective channel matrices $\effChannel{}$, the sequence of events that we envision is as follows:

\begin{enumerate}
\setlength{\itemsep}{1pt}
\setlength{\parskip}{1pt}

\item
At each update epoch $\dtime = 1,2\dotsc$, the user selects a transmit power profile $\power{\dtime} \in \state$.

\item
The user's loss $\loss{\power{\dtime}}{\dtime}$ is determined by the state of the network and the behavior of all other users via the effective channel matrices $\effChannel{\dtime}$ at the time of the user's transmission.

\item
The user selects a new transmit power profile $\power{\dtime+1} \in \state$ at stage $\dtime+1$ in an effort to minimize the \emph{a priori unknown} objective function $\loss{\power{}}{\dtime+1}$ and the process repeats.
\end{enumerate}

Needless to say, the key challenge in this dynamic framework is that the user does not know his objective function $\loss{\power{}}{\dtime}$ ahead of time, so he must try to somehow adapt to the changing network conditions ``on the fly'' (recall that $\loss{\power{}}{\dtime}$ depends at each stage $\dtime$ on the evolution of the environment and the choices of all other users).
As a result, static solution concepts (such as Nash or correlated equilibria) are no longer relevant because, in general, there is no optimum system state to target \textendash\ either static or in the mean.

Instead, given a time horizon $\timeHorizon$, we will compare the cumulative loss incurred by the user's power profile $\power{\dtime}$ for $\dtime = 1,2,\dotsc,\timeHorizon$, to the loss that the user would have incurred if he had chosen the \emph{best possible transmit profile in hindsight};
specifically, we define the user's \emph{regret} as:\begin{equation}
\label{eq:regret}
\Reg(\timeHorizon)
	= \max_{\fxPower{}\in\state}
	\insum_{\dtime=1}^{\timeHorizon} \big[ \loss{\power{\dtime}}{\dtime} - \loss{\fxPower{}}{\dtime} \big].
\end{equation}
The seminal notion of regret was first introduced in a game-theoretic setting by Hannan \cite{Han57} and it has since given rise to an extremely active field of research at the interface of optimization, statistics and machine learning \textendash\ for a recent survey, see e.g. \cite{CBL06,SS11}.%
\footnote{The terminology stems from the fact that large positive values of $\Reg(\timeHorizon)$ indicate that the user would have achieved a better power/rate trade-off in the past by employing some fixed $\fxPower{}$ instead of $\power{\dtime}$, making him ``regret'' his choice.}
The user's \emph{average regret} is then defined as $\timeHorizon^{-1} \Reg(\timeHorizon)$ and the goal of \emph{regret minimization} is to devise a dynamic transmit policy $\power{\dtime}$ which is asymptotically optimal in hindsight, i.e. that leads to \emph{no regret}:
\begin{equation}
\label{eq:no-reg}
\limsup\nolimits_{\timeHorizon\to\infty} \Reg(\timeHorizon) \big/ \timeHorizon
	\leq 0,
	\quad
	\text{or, equivalently:}
	\quad
\Reg(\timeHorizon)
	= o(\timeHorizon),
\end{equation}
\emph{irrespective} of how the objective function \eqref{eq:Loss} evolves over time.

\begin{remark}
Importantly, if the user's objective \eqref{eq:Loss} does not vary with time (or if it varies in a stochastic fashion, following some \ac{iid} process), a no-regret policy converges to the problem's static (or, respectively, average) solution \cite{SS11}.
Furthermore, if the user could predict the solution of \eqref{eq:OPC} ahead of every stage $\dtime=1,2,\dotsc,\timeHorizon$ in an oracle-like fashion, we would have $\Reg(\timeHorizon) \leq 0$ in \eqref{eq:regret} for all $\timeHorizon$;
by this token, the no-regret requirement \eqref{eq:no-reg} is an indicator that $\power{\dtime}$ tracks the optimum solution $\power{}^{\ast}(\dtime)$ of \eqref{eq:OPC} as it evolves over time.%
\footnote{In the machine learning literature, there exist more sophisticated notions of regret (such as adaptive \cite{HazSes09} or shifting \cite{CBGLS12} regret) that further quantify the quality of this tracking;
due to space limitations however, we will focus our theoretical analysis almost exclusively on external regret minimization which requires less technical language to describe.}
\end{remark}


\section{Adaptive Power Control via Exponential Learning}
\label{sec:PowerControlForMIMOOFDM}

In this section, we derive an adaptive \acl{PC} algorithm for the online optimization problem \eqref{eq:OPC} based on the method of \acf{MXL} \cite{TRW05,MBM12,KSST12}.
We first consider the case where the transmitter has access to perfect \acf{CSI};
the case of measurement errors and imperfect \ac{CSIT} is then discussed in Sec.~\ref{sec:PowerControlWithImperfectCSI}.

\subsection{Learning with perfect \ac{CSI}}
\label{sec:PowerControlWithPerfectCSI}

A key element in our approach will be the gradient $\gradient{} = \nabla_{\power{}}\, \lossPM{}{}$ of the user's objective function \eqref{eq:Loss}.
Specifically, if the rate-adjustment function $\phi$ is smooth,%
\footnote{In the general Lipschitz case, it suffices to replace $\phi'(R)$ by any element of $[\phi'(0^{-}),\phi'(0^{+})]$.}
we readily get:
\begin{equation}
\gradient{}
	= \nabla_{\power{}}\, \lossPM{}{}
	= \eye - \phi'(R) \cdot \nabla_{\power{}}\, R.
\end{equation}
Some matrix calculus then yields:
\begin{equation}
\nabla_{\power{}}\, R
	= \effChannelCT{} \inBra{\eye+\effChannel{} \power{} \effChannelCT{}}^{-1} \effChannel{},
\label{eq:RateDerivative}
\end{equation}
so the gradient of $\loss{\power{\dtime}}{\dtime}$ at $\power{\dtime}$ will be: 
\begin{equation}
\label{eq:LossDerivative}
\gradient{\dtime}
	= \eye - \phi'(\rate{\power{\dtime}}{\dtime}) \cdot
	\effChannelCT{\dtime}\,
	\big[
	\eye+\effChannel{\dtime}\,\power{\dtime}\,\effChannelCT{\dtime}
	\big]^{-1}\,
	\effChannel{\dtime}.
\end{equation}
Since the effective channel matrices $\effChannel{\dtime}$ are assumed bounded,
$\gradient{\dtime}$ will also be bounded for all $\dtime$;
hence, we formally assume that there exists a positive constant $\gradientB$ such that
\begin{equation}
\norm{\gradient{\dtime}}
	\leq \gradientB,
\label{eq:GradientUB}
\end{equation}
where $\norm{\gradient{}} = \lambda_{\max}(\gradient{})$ denotes the ordinary spectral norm (spectral radius) of $\gradient{}$.

In view of the above, a first idea would be to update the user's power profile $\power{\dtime}$ along the direction of steepest descent indicated by $\gradient{\dtime}$ \cite{Zin03};
however, this \acl{OGD} scheme would invariably violate the user's semidefiniteness constraint $\power{}\succcurlyeq 0$, so it is not a viable transmit policy.
Instead, inspired by the matrix regularization methods of \cite{TRW05,MBM12,KSST12}, we propose an algorithm that tracks the direction of steepest descent in a dual, unconstrained space and then maps the result back to the problem's state space via matrix exponentiation.
More precisely, assuming for the moment perfect \ac{CSIT}, we will consider the \acl{MXL} scheme:
\begin{equation}
\tag{MXL}
\label{eq:MXL}
\begin{aligned}
\cumGradient{\dtime}
	&= \cumGradient{\dtime-1} - \gradient{\dtime},
	\\
\power{\dtime+1}
	&= \pmax \frac{\exp( \etaSqrt{\dtime}\cumGradient{\dtime})}{1+\motr\big[ \exp(\etaSqrt{\dtime} \cumGradient{\dtime} \big]},
\end{aligned}
\end{equation}
where $\eta>0$ is a parameter that controls the user's learning rate and the recursion is initialized with $\cumGradient{0} = 0$.

\begin{algorithm}
  \SetKwInOut{Input}{input}
  \SetKwInOut{Parameter}{parameter}
  
  \Parameter{$\eta>0$}
  \BlankLine
	\tcc{Initialization}
  $\dtime\leftarrow 0$;
  $\cumGradient{} \leftarrow \bmat{0}$\;
  \BlankLine
  \Repeat{transmission ends}{
    $\dtime \leftarrow \dtime+1$\;
    \tcc{Pre-transmission: Set Power}
    $\displaystyle\power{} \leftarrow \pmax \frac{\exp(\etaSqrt{\dtime}\cumGradient{})}{1+\motr\big[ \exp(\etaSqrt{\dtime}\cumGradient{}) \big]}$\;
    \BlankLine
    \tcc{\textbf{Transmission}}
    \BlankLine
    \tcc{Post-transmission: Measure Rate and Effective Channel Matrices}
    $R \leftarrow \log\det \big( \eye+\effChannel{}\power{}\effChannelCT{} \big)$\;
    $\gradient{}
    	\leftarrow \eye
	- \phi'(R) \cdot \effChannelCT{} \inPar{\eye+\effChannel{} \power{} \effChannelCT{}}^{-1} \effChannel{}$\;
    $\cumGradient{} \leftarrow \cumGradient{} - \gradient{}$\;
	}
	\caption{Matrix exponential learning (\acs{MXL})}
	\label{alg:XLPC}
\end{algorithm}

The recursion \eqref{eq:MXL} will be the main focus of our paper, so some remarks are in order (for an algorithmic implementation, see Alg. \ref{alg:XLPC}):

\setcounter{remark}{0}

\begin{remark}
Intuitively, the exponentiation step in \eqref{eq:MXL} assigns more power to the spatial directions that perform well while the $\dtime^{-1/2}$ factor keeps the eigenvalues of $\power{\dtime}$ from approaching zero too fast (note that $\cumGradient{\dtime}$ grows as $\bigoh(\dtime)$);
the trace normalization then ensures that $\power{\dtime}$ satisfies the feasibility constraints of \eqref{eq:OPC} for all $\dtime\geq1$.
In particular, as we show in Appendix \ref{sec:ProofOfRegretBoundContinuousTime}, the recursion \eqref{eq:MXL} can be seen as a ``primal-dual'' \ac{OMD} method \cite{SS11} with a variable parameter \cite{KM14};
for an in-depth discussion, see \cite{SS11,MBM12,TRW05,KSST12,KM14} and references therein.
\end{remark}

\begin{remark}
From an implementation viewpoint, Algorithm~\ref{alg:XLPC} has the following desirable properties:
\begin{enumerate}
[(P1)]
\item
It is \emph{distributed}:
each transmitter updates his own power profile based only on local \ac{CSI}.
\item
It is \emph{asynchronous}:
the algorithm's updates are event-based and can be performed without synchronization or any further signaling/coordination between connections.
\item
It is \emph{agnostic}:
transmitters do not need to know the status or geographical distribution of other connections in the network.
\item
It is \emph{reinforcing}:
each connection tends to minimize its individual loss.
\end{enumerate}
\end{remark}

\begin{remark}
In terms of feedback, Algorithm \ref{alg:XLPC} requires that
\begin{inparaenum}
[\itshape a\upshape)]
\item
transmitters measure their achieved rates;
and
\item
the receiver feeds back to the transmitter the received signal covariance $\expectation{\rSgn{}\conjT{\rSgn{}}} = \interfPower{}+\channel{}\power{}\conjT{\channel{}}$ (e.g. via broadcasting or over a duplex downlink).
\end{inparaenum}
From a computational standpoint, it is then easy to see that the complexity of each iteration of Algorithm~\ref{alg:XLPC} is linear in the number of subcarriers $\nOfSubcarriers$ and polynomial in the number of transmit antennas $\nOfTx$:%
\footnote{We are implicitly assuming that $\phi'(R)$ can be calculated with very low cost \textendash\ e.g. by means of a lookup table.}
in particular, since $\cumGradient{}$ is block-diagonal, fast Coppersmith\textendash Winograd matrix multiplication \cite{DJ13} provides a worst-case complexity bound that is $\bigoh(\nOfSubcarriers\nOfTx^{2.373})$ per iteration.
\end{remark}

Our main theoretical result regarding the \ac{MXL} \acl{PC} algorithm (Alg.~\ref{alg:XLPC}) is as follows:
\begin{theorem}
\label{thm:RegretBound}
The \ac{MXL} algorithm \textup(Alg.~\ref{alg:XLPC}\textup) leads to no regret in the online \acl{PC} problem \eqref{eq:OPC}.
In particular, the iterates of \eqref{eq:MXL} enjoy the $\bigoh(\timeHorizon^{-1/2})$ regret bound:
\ifCLASSOPTIONdraftcls
\begin{equation}
\frac{1}{\timeHorizon}\Reg(\timeHorizon)
	\leq \inPar{\frac{\pmax\log\inPar{1+\nOfSubcarriers\nOfTx}}{\eta} + \frac{\eta\pmax\gradientB^2}{2}}\frac{1}{\sqrt{\timeHorizon}}
	+ \frac{\eta\pmax\gradientB^2}{4}\frac{1}{\timeHorizon},
\label{eq:RegretBound}
\end{equation}
	\else
\begin{equation}
\begin{aligned}
\frac{1}{\timeHorizon} \Reg(\timeHorizon)
	& \leq \inPar{\frac{\pmax\log\inPar{1+\nOfSubcarriers\nOfTx}}{\eta} + \frac{\eta\pmax\gradientB^2}{2}}\frac{1}{\sqrt{\timeHorizon}}
	\\
	& \quad + \frac{\eta\pmax\gradientB^2}{4}\frac{1}{\timeHorizon},
\end{aligned}
\label{eq:RegretBound}
\end{equation}
\fi
irrespective of how the system evolves over time.
\end{theorem}

\begin{IEEEproof}
  See Appendix \ref{sec:ProofOfRegretBoundWithPerfectCSI}.
\end{IEEEproof}

The bound \eqref{eq:RegretBound} is our main performance guarantee for Alg.~\ref{alg:XLPC}, so we proceed with a few remarks:

\begin{remark}
Even though Theorem \ref{thm:RegretBound} focuses on a given connection $\connection\in\connections$, the focal connection is still subject to interference from other connections in the network
(the incurred interference is captured by the effective channel matrices $\effChannel[\subcarrier]{}$ which depend on the interfering users' transmit policies).
In this light, Theorem \ref{thm:RegretBound} provides a worst-case performance guarantee which holds even in the presence of malicious users (jammers) that seek to shut down the focal connection.

On the other hand, a natural question that arises is whether users can meet more sophisticated criteria (such as reaching a globally efficient state or a Nash equilibrium) when they all follow the same algorithm and the wireless medium is otherwise static.
In the \ac{MIMO} \acl{MAC} (where all users transmit to a common receiver), it can be shown that the \ac{MXL} algorithm leads to a socially optimum state;
a more general treatment of this question (e.g. in the \ac{MIMO} interference channel \cite{SPB09-sp}) lies beyond the scope of this paper, so we delegate it to future work.
\end{remark}

\begin{remark}
We should also note here that the first term of the bound \eqref{eq:RegretBound} captures the dimensionality of the problem while the rest is an increasing function of the channel variability estimate $\gradientB$;
as such, the learning parameter $\eta$ of Algorithm~\ref{alg:XLPC} can be fine-tuned to accelerate the algorithm's convergence to a no-regret state in terms of $\gradientB$.
Specifically, the value of $\eta$ which minimizes the dominant $\bigoh(\timeHorizon^{-1/2})$ term of the regret bound \eqref{eq:RegretBound} for a fixed time horizon $\timeHorizon$ is:
\begin{equation}
\eta
	= \gradientB^{-1} \sqrt{2\log\inPar{1+\nOfSubcarriers\nOfTx}}.
\end{equation}
In turn, this parameter choice leads to the optimized convergence rate:
\begin{equation}
\label{eq:OptimalRegretBound}
\frac{1}{\timeHorizon} \Reg(\timeHorizon)
	\leq \pmax\gradientB \sqrt{2\log\inPar{1+\nOfSubcarriers\nOfTx}}
	\inPar{\frac{1}{\sqrt{\timeHorizon}} + \frac{1}{4\timeHorizon}}.
\end{equation}
The $\bigoh(1/\sqrt{\timeHorizon})$ dependence of \eqref{eq:OptimalRegretBound} is known to be asymptotically tight in the context of online optimization problems against an adversarial nature \cite{SS11},
while the $\bigoh(\log \nOfSubcarriers \nOfTx)$ behavior represents a significant reduction in the dimensionality of the problem (which has $\bigoh(\nOfSubcarriers \nOfTx)$ degrees of freedom).
In fact, \eqref{eq:OptimalRegretBound} becomes tight only in adversarial environments (e.g. induced by jamming), so, in practical situations, the user's regret minimization rate is considerably faster \textendash\ cf. Section \ref{sec:NumericalResults}.
\end{remark}

\begin{remark}
The agnostic initialization $\cumGradient{0} = 0$ is a conservative choice reflecting the worst-case scenario where the user assumes bad channel conditions.
Indeed, $\cumGradient{0}=0$ corresponds to initial transmit power equal to $\pmax \cdot \nOfSubcarriers\nOfTx/(1+\nOfSubcarriers\nOfTx) \sim \pmax$ in the large $\nOfSubcarriers$ (or large $\nOfTx$) limit;
in this way,
the user's transmit power will likely be reduced under Algorithm \ref{alg:XLPC} in the presence of good channel conditions.
Hence, if the transmitter has some estimate of his expected channel conditions, it would be preferable to initialize power accordingly:
if the user expects a good channel, initial power should be set lower (to save battery life);
otherwise, if a bad channel is expected, initial transmit power should be set high so as to avoid very low transmission rates in the first few frames.
\end{remark}

\subsection{Adaptive \acl{PC} with imperfect \ac{CSI}}
\label{sec:PowerControlWithImperfectCSI}

In practice, a major challenge occurs if the transmitters do not have access to perfect \ac{CSI} with which to update the adaptive \acl{PC} scheme \eqref{eq:MXL}.
In particular, given that each user's gradient matrix $\gradient{}$ is determined by his effective channel matrix $\effChannel{}$, imperfect measurements of the users' channel or the multi-user interference-plus-noise (due e.g. to pilot contamination, undersampling or other factors) could have a catastrophic effect on the no-regret properties of the proposed scheme \eqref{eq:MXL}.
Accordingly, our goal in this section will be to examine the robustness of \eqref{eq:MXL} in the presence of measurement errors and observation noise.


To model errors of this kind, we assume that, at each update period $\dtime =1,2,\dotsc$, the transmitter observes a noisy estimate $\gradientN{\dtime}$ of the form
\begin{equation}
\gradientN{\dtime}
	= \gradient{\dtime} + \noiseGradient{\dtime},
\label{eq:NoisyGradient}
\end{equation}
where the error process $\noiseGradient{\dtime} = \diag{\noiseGradient[1]{\dtime},\ldots,\noiseGradient[\nOfSubcarriers]{\dtime}}$ satisfies the statistical hypotheses:
\begin{enumerate}
[({H}1)]
\item
\label{hyp:zeromean}
\emph{Unbiasedness:}
\begin{equation}
\tag{H1}
\label{eq:zeromean}
\ex\big[
	\noiseGradient{\dtime} \,\vert\, \power{\dtime-1}
	\big]
	= 0.
\end{equation}

\item
\label{hyp:tailbound}
\emph{Tame tails:}
\begin{equation}
\tag{H2}
\label{eq:tailbound}
\prob\big( \norm{\noiseGradient{\dtime}} \geq z \big)
	\leq A/z^{\alpha}
	\quad
	\text{for some $A>0$ and for some $\alpha>4$.}
\end{equation}
\end{enumerate}
%

The unbiasedness assumption \eqref{eq:zeromean} is a bare-bones assumption which simply boils down to asking that there is no biased, \emph{systematic} error in the user's \ac{CSI} measurements.
Likewise, \eqref{eq:tailbound} posits a fairly mild control on the probability of observing very high errors, and is satisfied by the vast majority of statistical error distributions (including for instance uniformly distributed, Gaussian, log-normal, Weibull and Lévy-type error processes);
in particular, \emph{we do not assume} that the measurement errors $\noiseGradient{\dtime}$ are \ac{iid}, state-independent, or even a.s. bounded.

Importantly, under these mild hypotheses for the statistics of the measurement noise, we have:

\begin{theorem}
\label{thm:ExpectedRegretBound}
%
%
The \ac{MXL} algorithm \textup(Alg.~\ref{alg:XLPC}\textup) run with imperfect observations satisfying \eqref{eq:zeromean} and \eqref{eq:tailbound} leads to no regret \textup(a.s.\textup);
in particular, it enjoys the mean regret bound:
\ifCLASSOPTIONdraftcls
\begin{equation}
\expectation{\frac{1}{\timeHorizon} \Reg(\timeHorizon)} 
	\leq \inPar{\frac{\pmax\log\inPar{1+\nOfSubcarriers\nOfTx}}{\eta} + \frac{\eta\pmax\bar\gradientB^2}{2}}\frac{1}{\sqrt{\timeHorizon}} 
    + \frac{\eta\pmax\bar\gradientB^{2}}{4}\frac{1}{\timeHorizon},
\label{eq:ExpectedRegretBound}
\end{equation}
\else
  \begin{equation}
  \begin{aligned}
  \expectation{\frac{1}{\timeHorizon} \Reg(\timeHorizon)} & \leq \inPar{\frac{\pmax\log\inPar{1+\nOfSubcarriers\nOfTx}}{\eta} + \frac{\eta\pmax\gradientB^2}{2}}\frac{1}{\sqrt{\timeHorizon}}\\
	  & \quad + \frac{\eta\pmax\gradientB^2}{4}\frac{1}{\timeHorizon},
  \end{aligned}
  \label{eq:ExpectedRegretBound}
  \end{equation}
\fi
where $\bar\gradientB^{2} = \sup_{\dtime} \ex\big[ \smallnorm{\gradientN{\dtime}}^{2} \,\vert\, \power{\dtime-1} \big]^{2}$.
\end{theorem}

\begin{IEEEproof}
  See Appendix \ref{sec:ProofOfRegretBoundWithImperfectCSI}.
\end{IEEEproof}

\begin{remark}
From an implementation perspective, we should note here that the mean bound \eqref{eq:ExpectedRegretBound} reduces to the deterministic bound \eqref{eq:RegretBound} in the case of perfect \ac{CSI}.
Also, even though we have $\limsup_{\timeHorizon\to\infty} \timeHorizon^{-1} \Reg(\timeHorizon) \leq 0$ (a.s.), the realized regret of Alg.~\ref{alg:XLPC} may exceed the mean bound \eqref{eq:RegretBound} with positive probability.
By a concentration inequality argument \cite{HH80}, it is possible to estimate analytically the probability of such deviations in terms of the central moments of the error process, but this analysis would take us too far afield so we do not present it here.
\end{remark}

\begin{remark}
Hypothesis \eqref{eq:tailbound} implies that the error process $\noiseGradient{}$ has finite (central) moments of up to fourth order \textendash\ in fact, barring pathological examples, this requirement is essentially tantamount to \eqref{eq:tailbound}.
The importance of fourth order moments has to do with the fact that we are using a variable learning parameter that decays as $\dtime^{-1/2}$;
by choosing a slower decay rate of the form $\dtime^{-\gamma}$ for some $\gamma\in(0,1/2)$, it is possible to relax Hypothesis \eqref{eq:tailbound} down to second order moment control.
However, given that \eqref{eq:tailbound} already suffices for the framework at hand (and due to space limitations), we do not present this more general analysis here.
\end{remark}

\section{Numerical Results}
\label{sec:NumericalResults}

\acused{OFDM}

To validate the theoretical analysis of Section \ref{sec:PowerControlForMIMOOFDM}, we conducted extensive numerical simulations over a wide range of design parameters and specifications.
In what follows, we present a representative subset of these results, but the conclusions drawn remain valid in most typical mobile wireless environments.

Throughout this section, we consider a typical cellular \ac{OFDMA} wireless network that occupies a $10\,\mathrm{MHz}$ band divided into $1024$ subcarriers around a central frequency $f_{c} = 2.5\,\mathrm{GHz}$.
We further assume that each cell employs a simple randomized access algorithm \cite{stiakogiannakis_radio_2013} to allocate subcarriers to the users it serves.
In the following, we focus on $\nOfConnections = 4$ users that are located at different cells \textendash\ served by different \acp{BS} \textendash\ and that have been allocated the same set of $\nOfSubcarriers = 8$ subcarriers. 
We focus on the \ac{UL} case, so the receivers are assumed stationary whereas the transmitters may be either stationary or mobile, depending on the simulated scenario.
Communication occurs over a \ac{TDD} scheme with frame duration $T_{f} = 5\,\mathrm{ms}$:
specifically, transmission occurs during the \ac{UL} subframe while receivers process the transmitted signal and provide feedback during the \ac{DL} subframe;
upon reception of the feedback, transmitters update their transmit powers according to Algorithm \ref{alg:XLPC}, and the process repeats until transmission ends.
For demonstration purposes, we simulated the case where each connection has a fixed rate requirement $\tRate[\connection]{}$ which varies across connections $\connection\in\connections$ so as to ensure diversity of \ac{QoS} requirements (the users' tolerance and loss function is defined as indicated in Remark \ref{rem:rate-req}).
For convenience, all simulation parameters are summarized in Table \ref{tab:OFDMANetworkSimulationParameters}.

\begin{table}[t!]
	\caption{OFDMA Network Simulation Parameters}
	\label{tab:OFDMANetworkSimulationParameters}
	\centering
	\begin{tabularx}{\linewidth}{XX}
		\hline
		\hline
		Number of Cells 		& $19$\\
		Cell Radius 			& $1\,\mathrm{km}$\\
		Central Frequency 		& $2.5\,\mathrm{GHz}$\\
		Available Bandwidth 		& $10\,\mathrm{MHz}$\\
		Number of \ac{OFDM} Subcarriers 	& $1024$\\
		Subcarrier Spacing		& $10.9375\,\mathrm{kHz}$\\
		Tranmit Antennas		& $2$\\
		Receive Antennas		& $2$\\
		Propagation Model 		& COST-Hata-Model \\ 
		BS Antenna Height 		& $32\,\mathrm{m}$\\
		MS Antenna Height 		& $1.5\,\mathrm{m}$\\
		Shadowing 			& $8.9\,\mathrm{dB}$\\
		AWGN Spectral Power Density 	& $N_0=-174\;\mathrm{dBm}/\mathrm{Hz}$\\
		Receiver Noise Figure 		& $7\,\mathrm{dB}$\\
		Frame duration			& $5\,\mathrm{ms}$ \\
		
		Requested Bit Rate per User 	& $\{764.6, 113.7, 909.3, 1081.3\}\,\mathrm{kbps}$\\
		Maximum transmit power per User	& $\{40.40, 41.10, 42.85, 45.58\}\,\mathrm{dBm}$\\
	\hline
	\end{tabularx}
\end{table}

\begin{figure}[t]
\centering
\subfloat[Loss]{\includegraphics[width=\figWidthTwo]{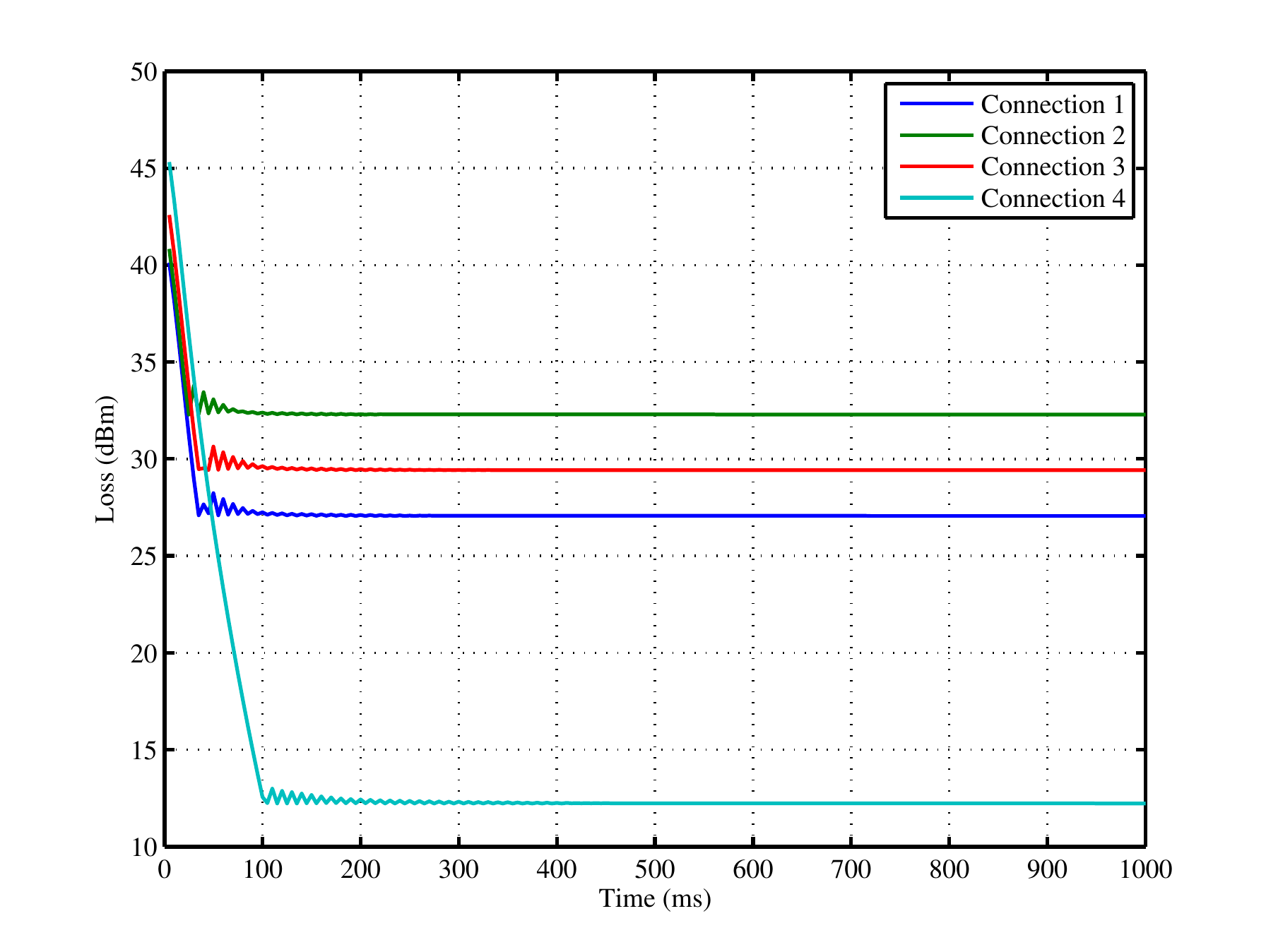}\label{fig:static:loss}}%
\hfill
\subfloat[Power]{\includegraphics[width=\figWidthTwo]{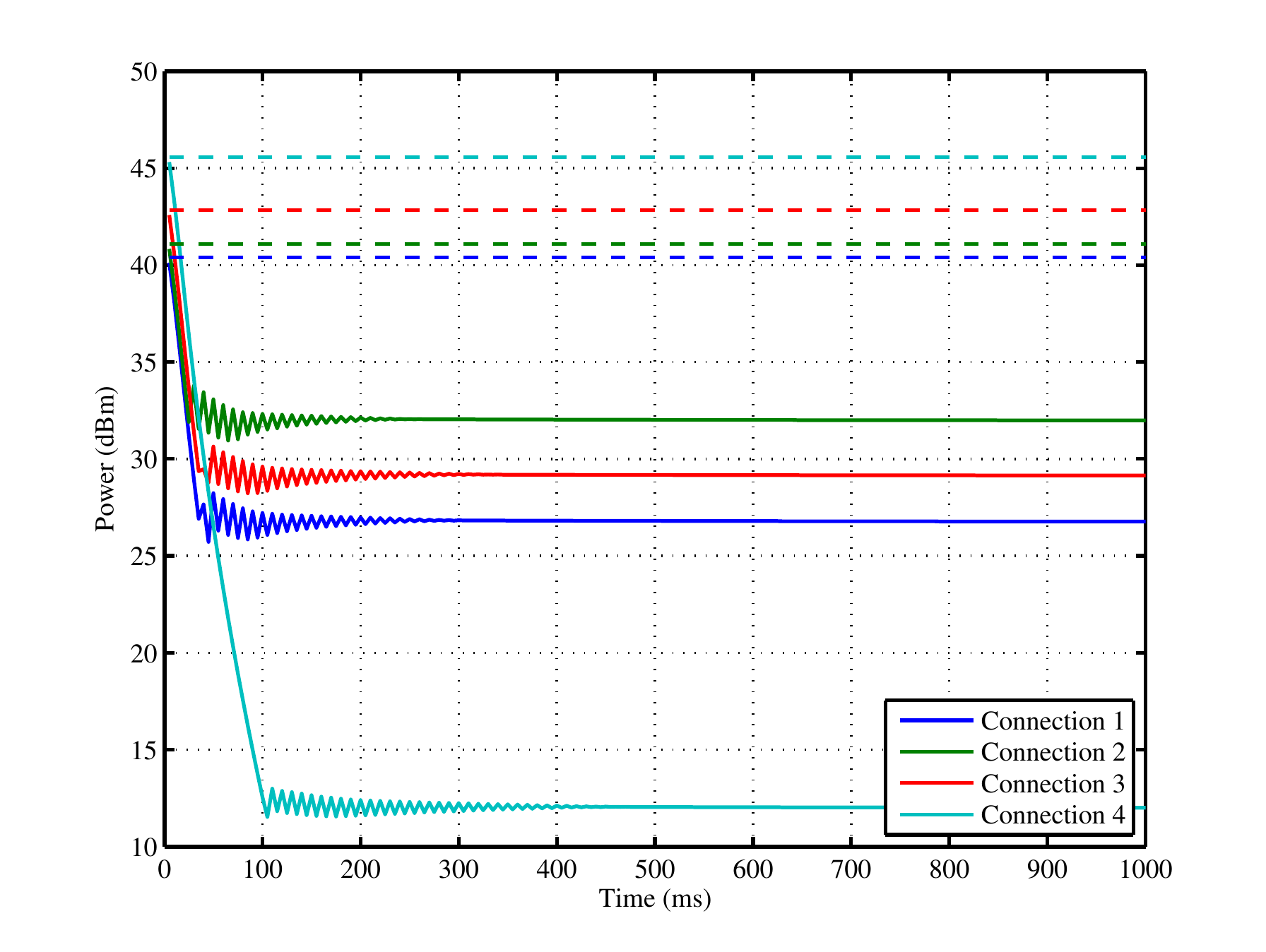}\label{fig:static:power}}\\%
\subfloat[Rate]{\includegraphics[width=\figWidthTwo]{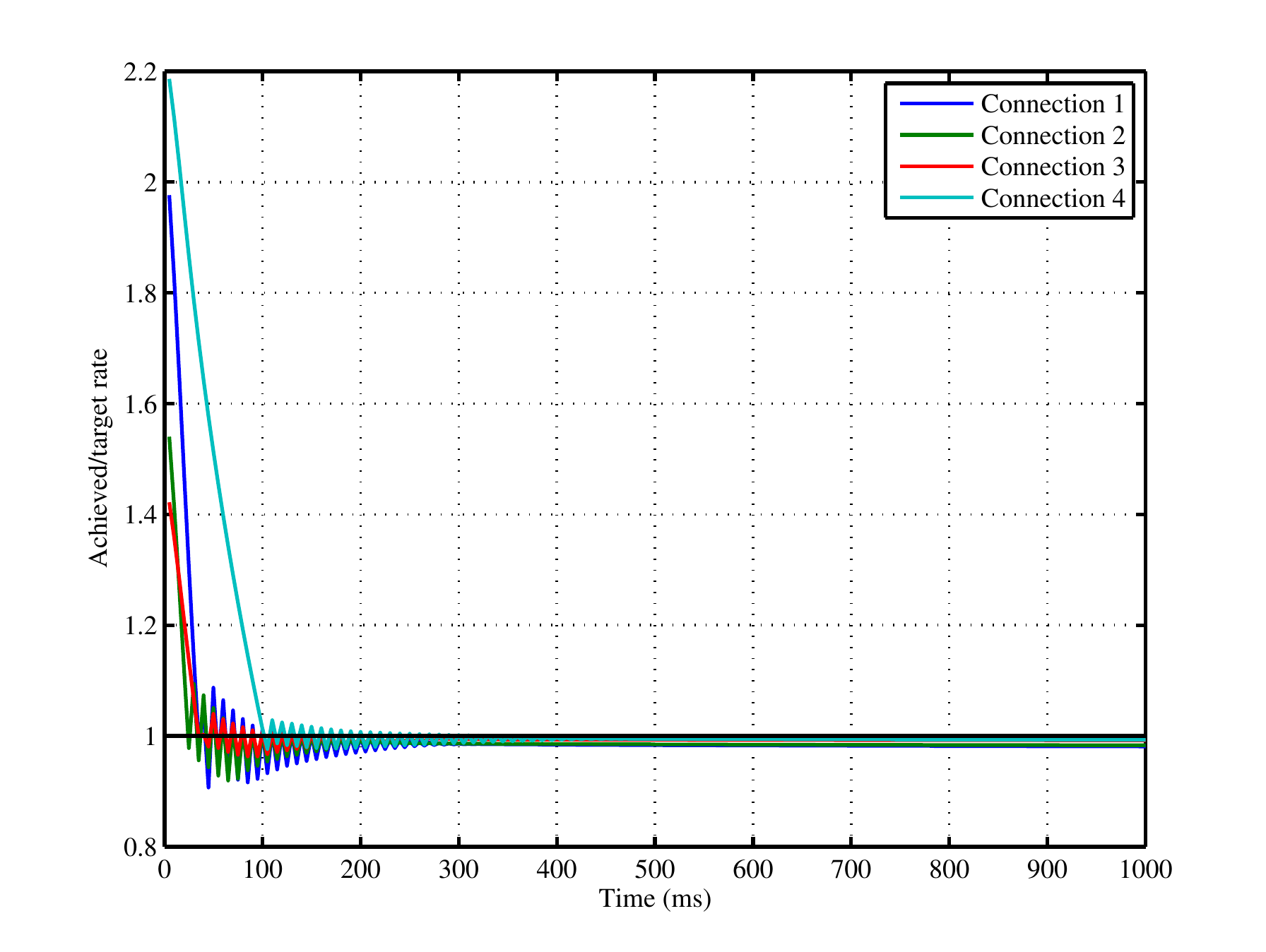}\label{fig:static:rate}}%
\hfill
\subfloat[Average Regret]{\includegraphics[width=\figWidthTwo]{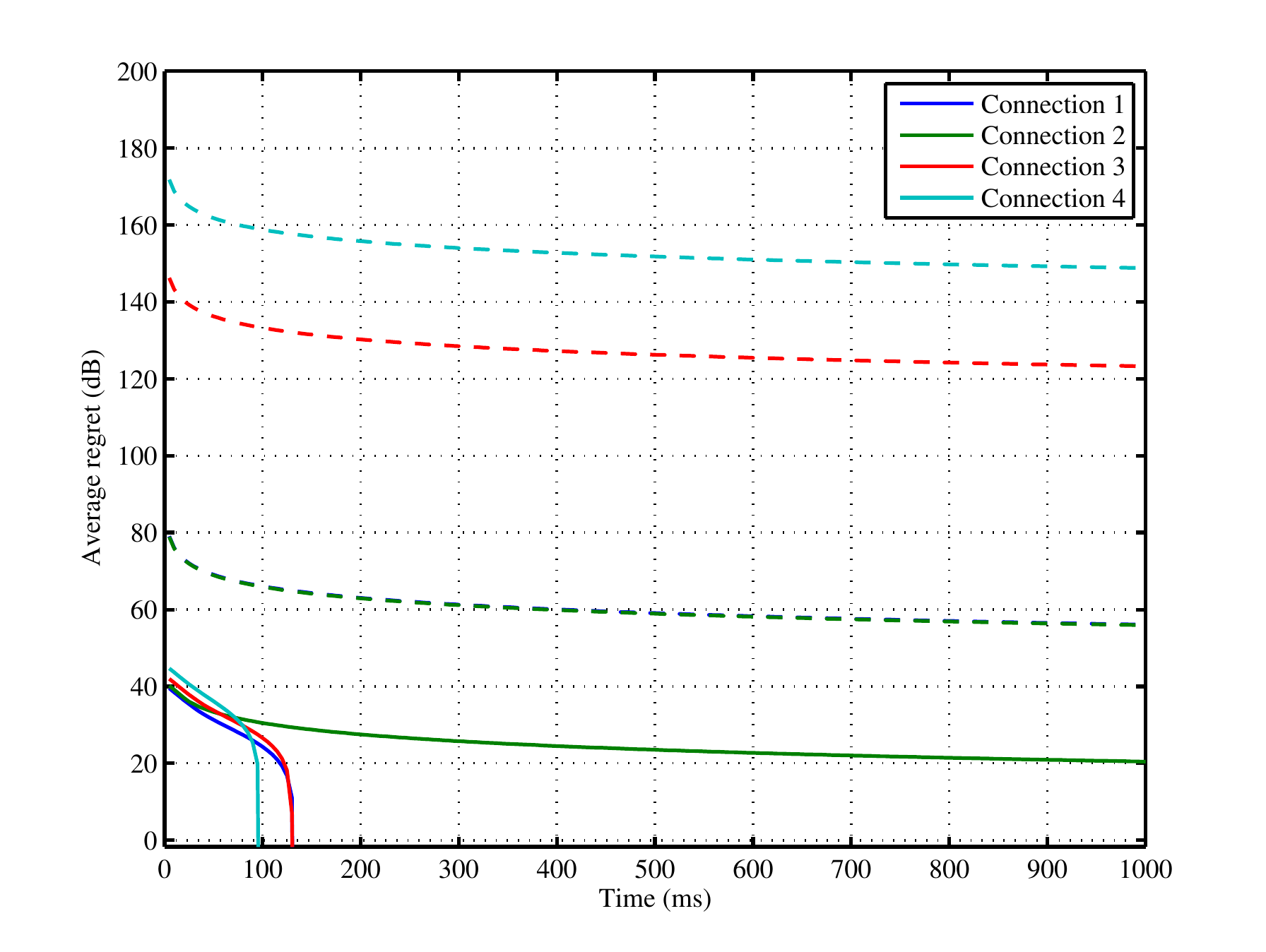}\label{fig:static:regret}}%
%
\caption{%
Adaptive \ac{MIMO}\textendash\ac{OFDM} power control under total power constraints for $\nOfConnections=4$ connections with static channel conditions. %
Fig. \ref{fig:static:loss} depicts the evolution of the users' objective function $\loss{\power{}}{\dtime}$ under the online power control algorithm \eqref{eq:MXL}.
Similarly, Fig. \ref{fig:static:power} shows the evolution of the users' total transmit power $\tr{\power{\dtime}}$ (solid lines; dashed lines correspond to the users' maximum transmit power);
Fig. \ref{fig:static:rate} shows the achieved/target rate gap $r\inPar{\power{};\dtime}  \big/ \tRate{}$.
Finally, the average regret $\dtime^{-1} \regret{\fxPower}{\dtime}$ is plotted in Fig. \ref{fig:static:regret} (solid lines), along with the theoretical bounds predicted by Theorem \ref{thm:RegretBound} (dashed lines); for simplicity, we only plot the positive part in the regret and we use a logarithmic dB scale for consistency.
}
\label{fig:static}
\end{figure}

For benchmarking purposes, the first simulated scenario focuses on the case where channels remain static during the transmission horizon.
In Fig. \eqref{fig:static:loss}, we plot the evolution of the users' objective $\loss{\power{}}{\dtime}$ under Algorithm \ref{alg:XLPC}:
as can be seen, users quickly reach an optimal state corresponding to the minimum of their loss function (i.e. minimum transmit power subject to the users' rate requirements).
In particular, as we see in Fig.~\eqref{fig:static:power}, even though all connections start with excessive transmit power (due to the algorithm's conservative initialization), they converge within $2\,\mathrm{dB}$ of their optimum transmit profile within a few frames (between $5$ and $15$, depending on the connection).
Interestingly, we also see some slight power oscillations (of the order of $1\,\mathrm{dB}$) that persist for a few frames after the initial ones:
these are due to small violations of the users' rate requirements (due to the power updates of other users) that cause them to momentarily increase their transmit power.
%
%
Similar oscillations are observed with respect to the achieved/target rate gap $\rate{\power{}}{\dtime} \big/ \tRate{}$ depicted in Fig.~\eqref{fig:static:rate}:
users quickly get within $2\textendash5\%$ of their target value, but they oscillate slightly for a few frames before converging.
%
Finally, in Fig.~\eqref{fig:static:regret}, we plot the user's average regret $\timeHorizon^{-1}\Reg(\timeHorizon)$ (solid lines) along with the theoretical bound predicted by Theorem \ref{thm:RegretBound} (dashed lines).
To increase resolution, we plot the users' regret in a logarithmic scale:
in this way, the observed vertical drops to $-\infty$ correspond to the point where the users' regret becomes negative (an indication of the number of frames required for the algorithm to converge).
In tune with the above observations, we see that users only require a few frames to achieve a no-regret state.

\begin{figure}[t]
\centering
\subfloat[Power]{\includegraphics[width=\figWidthTwo]{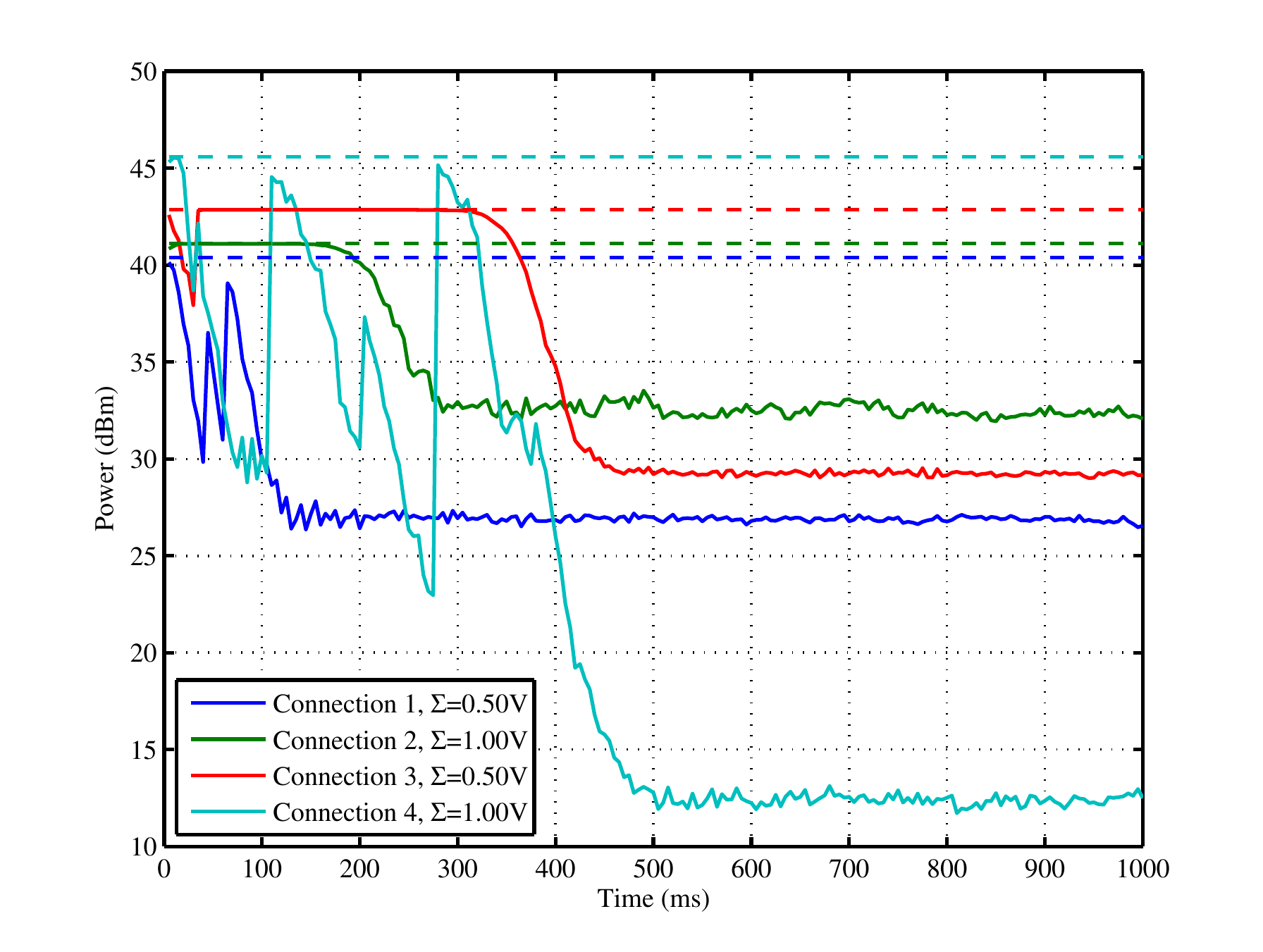}\label{fig:noisyStatic:power}}%
\hfill
\subfloat[Rate]{\includegraphics[width=\figWidthTwo]{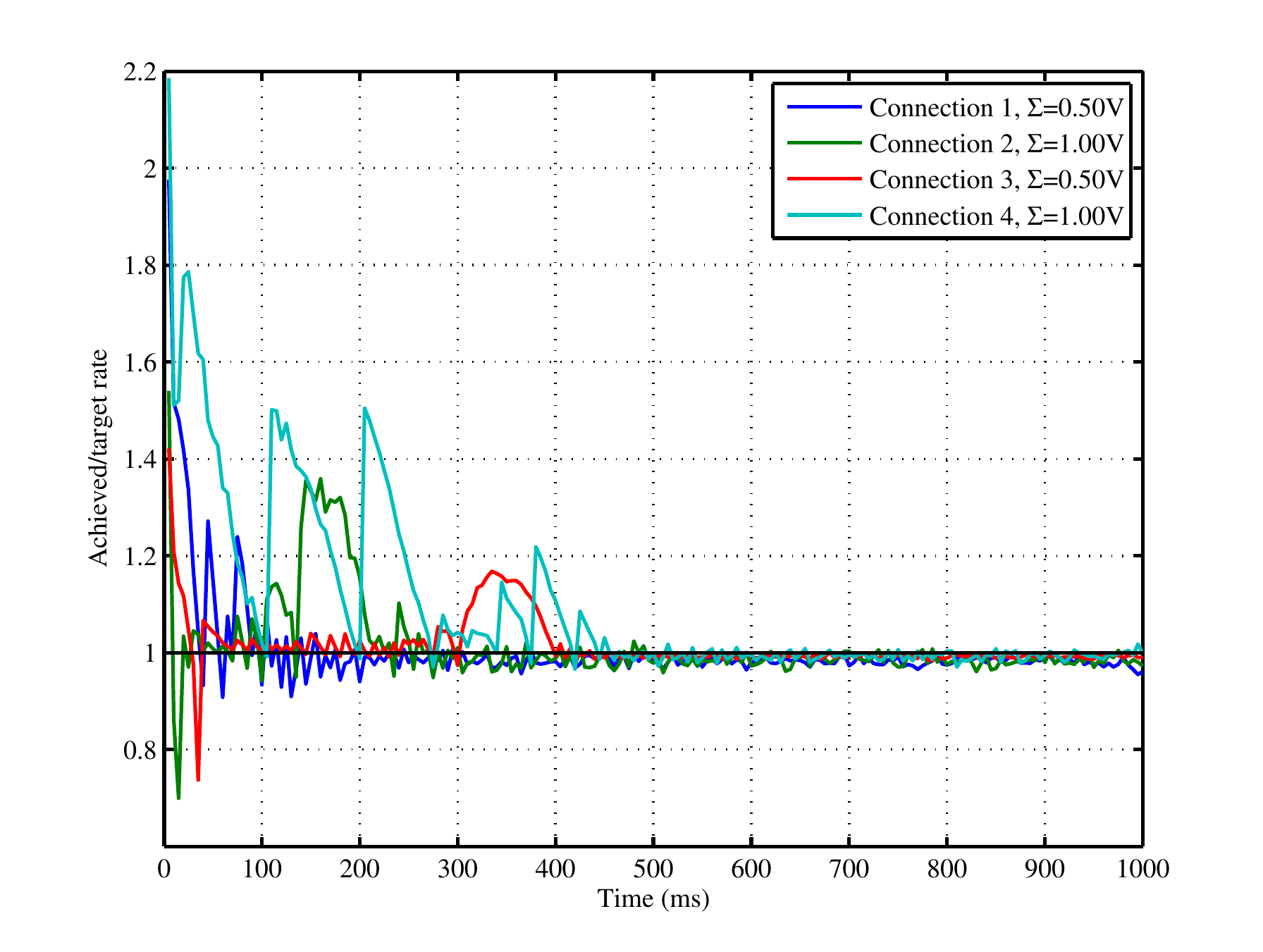}\label{fig:noisyStatic:rate}}\\%
\subfloat[Average Regret]{\includegraphics[width=\figWidthTwo]{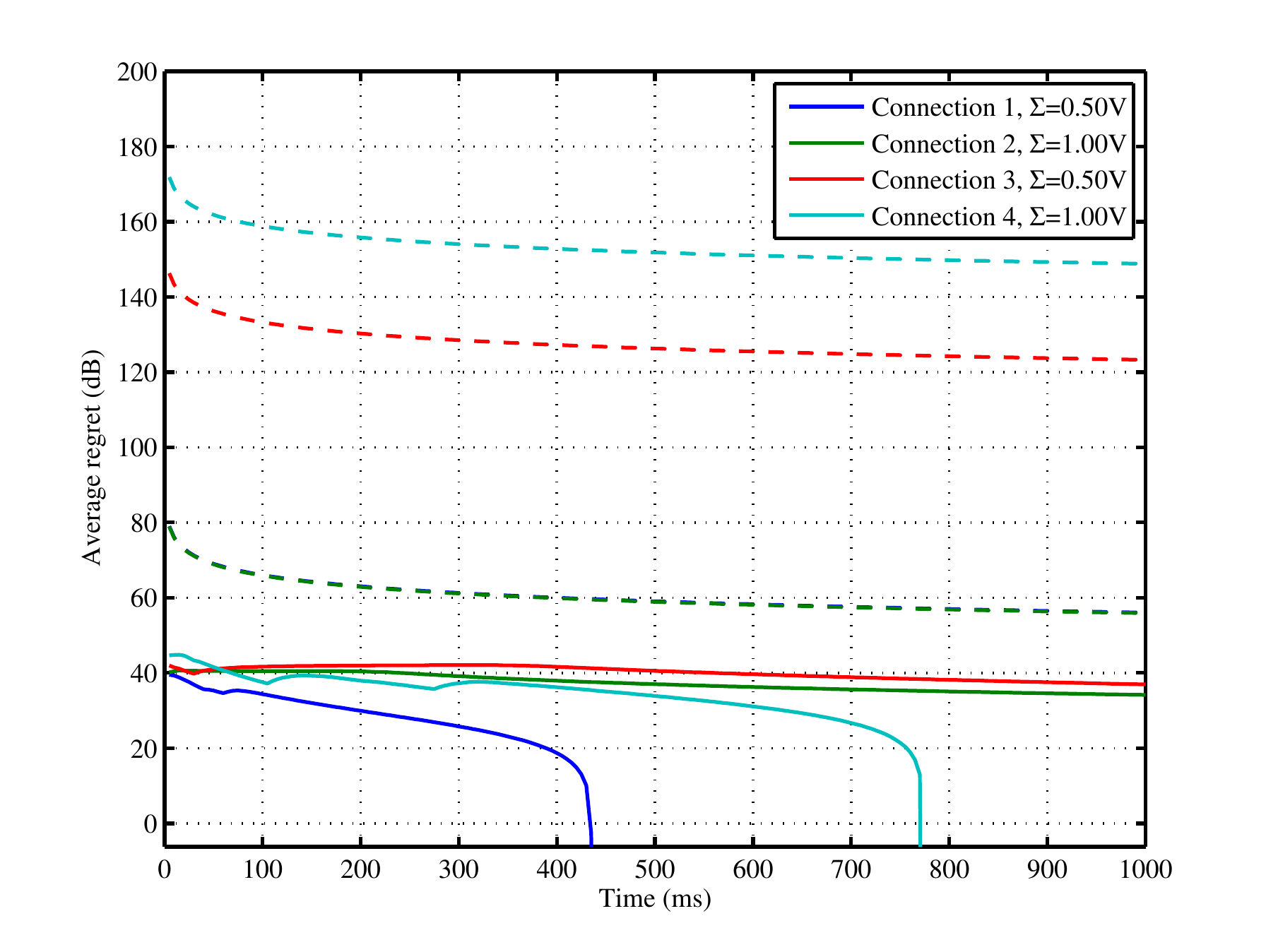}\label{fig:noisyStatic:regret}}%
%
\caption{
Adaptive \ac{MIMO}\textendash\ac{OFDM} power control for $\nOfConnections=4$ connections with static channel conditions but imperfect \ac{CSIT} (relative measurement errors depicted in each figure's legend).
Figs. \ref{fig:noisyStatic:power} and \ref{fig:noisyStatic:rate} respectively depict the evolution of the users' radiated power $\tr{\power{\dtime}}$ and achieved/target rate gap $r\inPar{\power{};\dtime} \big/ \tRate{}$ under the online \acl{PC} algorithm \eqref{eq:MXL}.
The users' average regret $\dtime^{-1} \regret{\fxPower}{\dtime}$ is plotted in Fig. \ref{fig:noisyStatic:regret} (solid lines), along with the theoretical bounds predicted by Theorem \ref{thm:RegretBound} (dashed lines);
for simplicity, we only plot the positive part of the regret and we use a logarithmic dB scale for consistency.}
\label{fig:noisyStatic}
\end{figure}

The second simulated scenario examines the case of imperfect \ac{CSI}.
Specifically, in Fig.~\ref{fig:noisyStatic}, we consider the same network realization as in Fig.~\ref{fig:static}, but we no longer assume that transmitters receive perfect \ac{CSI} during the \ac{TDD} feedback loop;
instead, we assume imperfect channel state measurements and we plot the users' power, rate and regret under Algorithm \ref{alg:XLPC} with noisy observations.
In particular, the transmitters' \ac{CSI} deviates from its corresponding mean value with standard deviation of $0.50\gradientB$ for connections 1 and 3, and $1.00\gradientB$ for connections 2 and 4, respectively.
Due to this huge uncertainty,
users are more conservative and tend to use up more power to achieve their rate requirements;
however, after an initial sampling period (lasting a few tens of frames), they confidently reduce power and converge to an optimum rate/power trade-off (as evidenced by the minimization of their objecitve).
A similar behavior is observed in Fig.~\eqref{fig:noisyStatic:rate} which shows the evolution of the users' throughput over time:
even though there are more pronounced fluctuations over the first few frames, all connections eventually converge to their target rates.
%
The main performance degradation is in the algorithm's convergence time:
as can be seen in Fig.~\ref{fig:noisyStatic:regret}, Algorithm \ref{alg:XLPC} takes longer to converge to a no-regret state, chiefly due to the regret generated during the algorithm's training phase.

\begin{figure}[t]
\centering
\subfloat[Channel]{\includegraphics[width=\figWidthTwo]{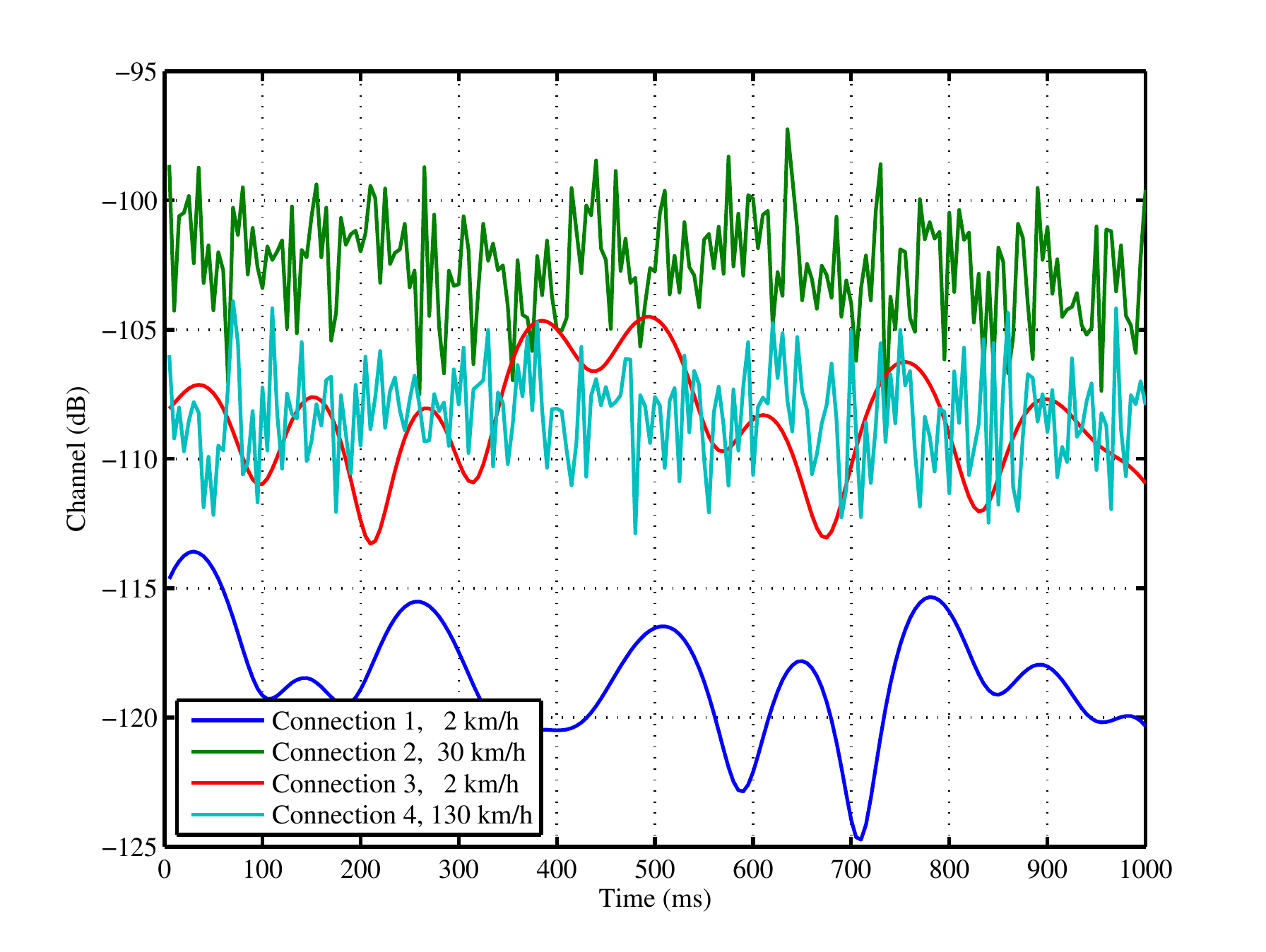}\label{fig:timeVarying:channel}}%
\hfill
\subfloat[Power]{\includegraphics[width=\figWidthTwo]{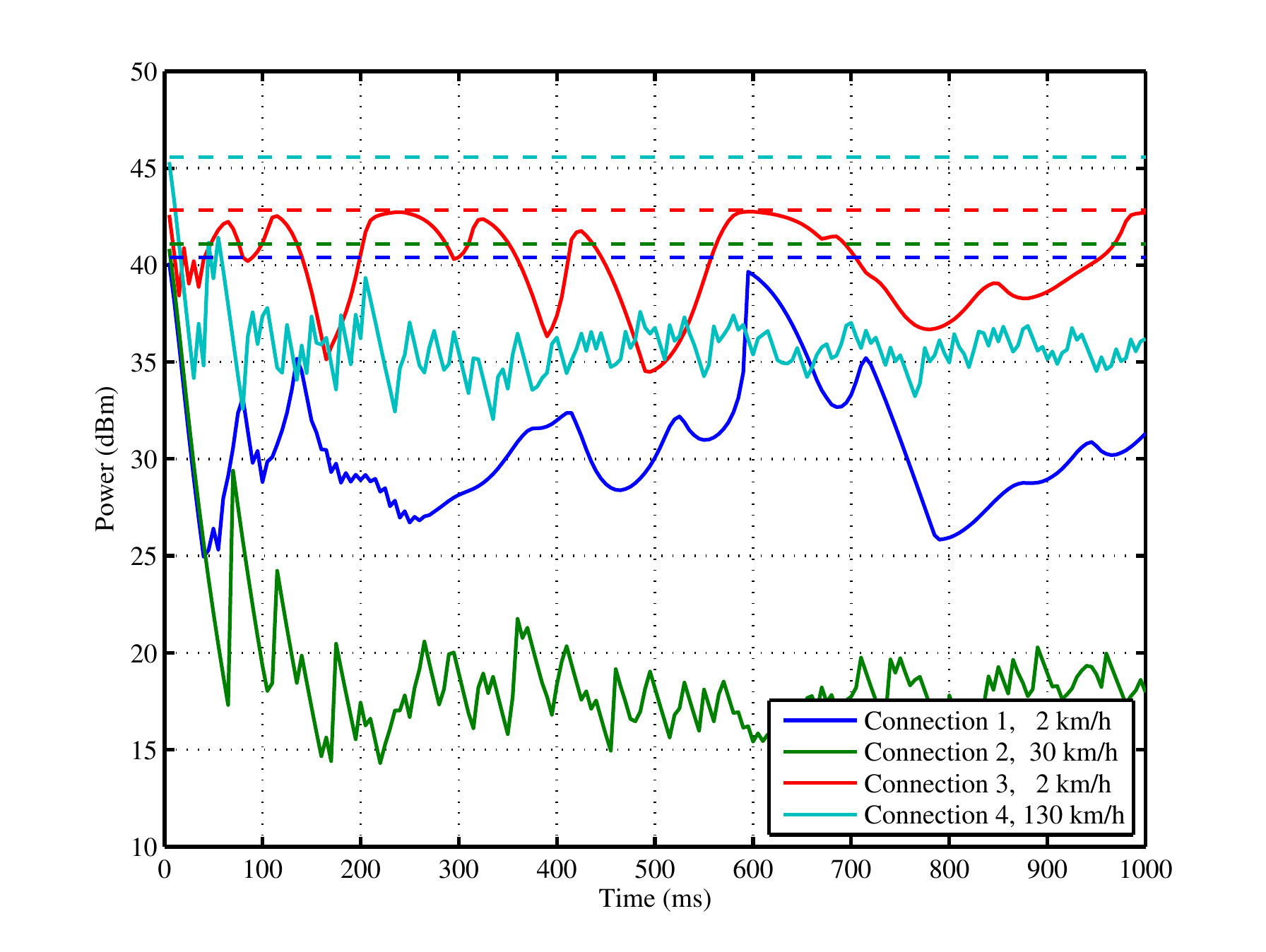}\label{fig:timeVarying:power}}\\%
\subfloat[Average Rate]{\includegraphics[width=\figWidthTwo]{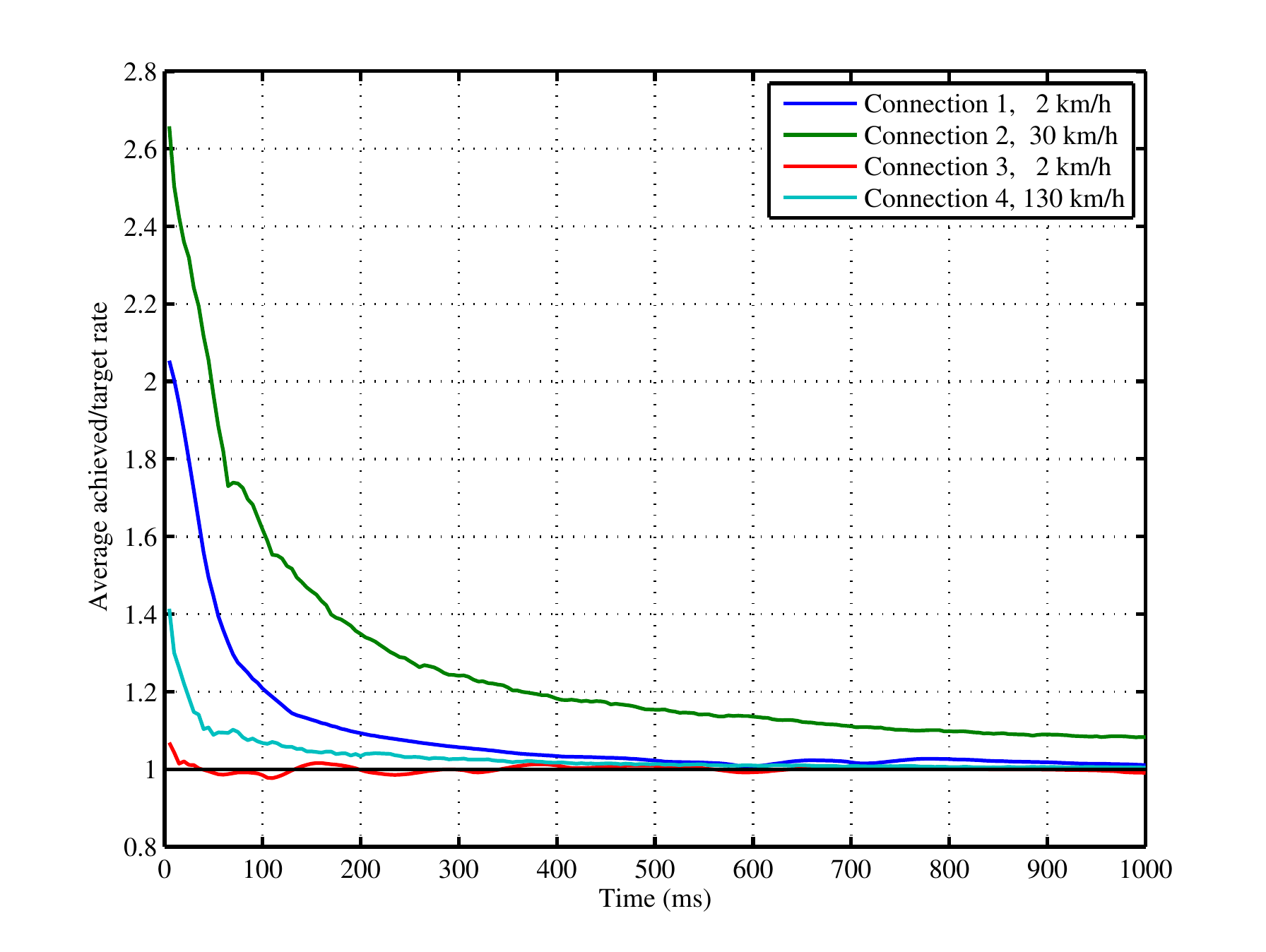}\label{fig:timeVarying:averageRate}}%
\hfill
\subfloat[Average Regret]{\includegraphics[width=\figWidthTwo]{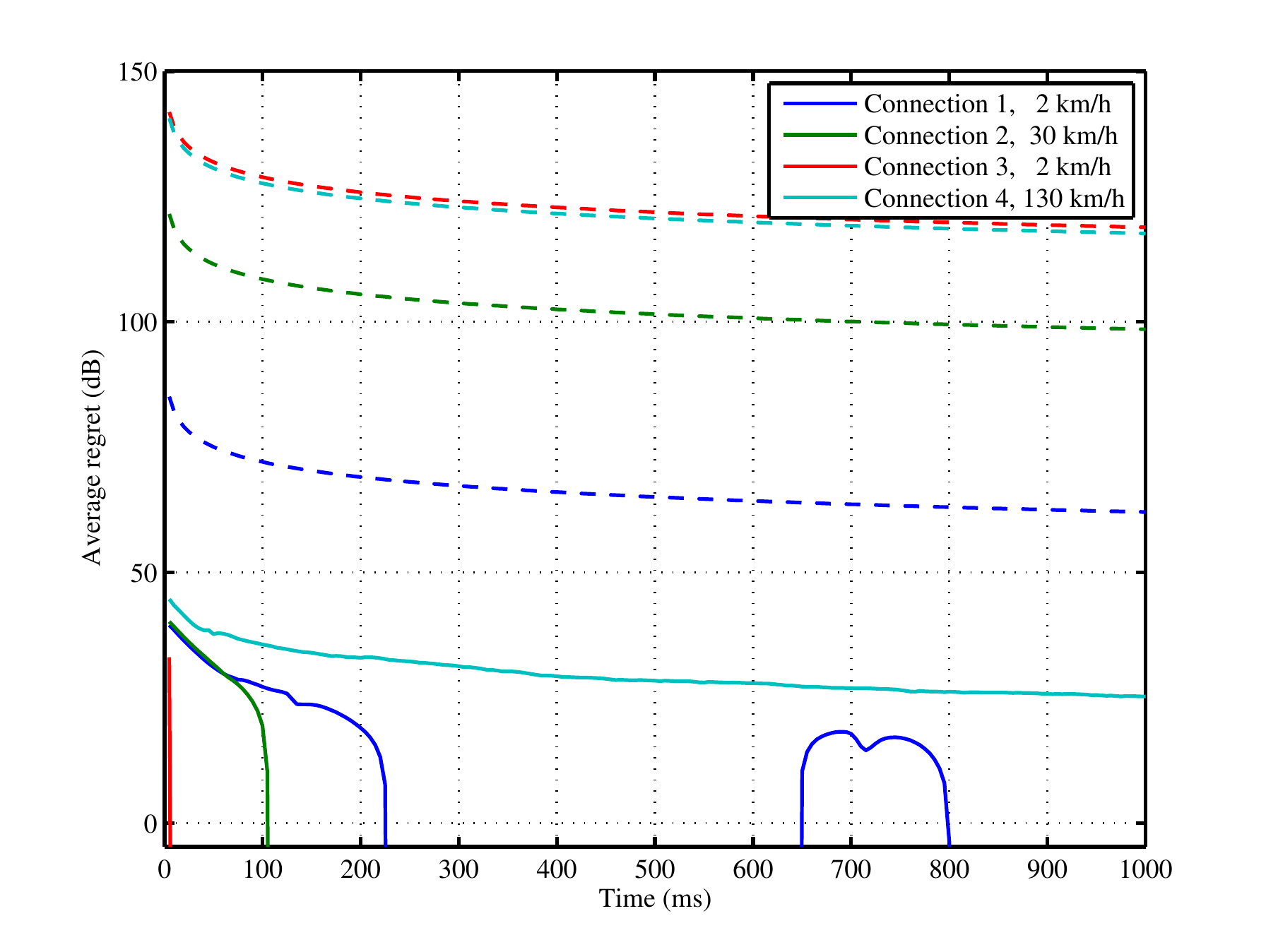}\label{fig:timeVarying:regret}}%
%
\caption{
Adaptive \ac{MIMO}\textendash\ac{OFDM} power control for $\nOfConnections=4$ connections with time-varying channel conditions corresponding to stationary receivers and mobile transmitters (average speed as in each figure's legend).
For reference purposes, Fig.~\ref{fig:timeVarying:channel} depicts the evolution of the channel gains $\motr\big[ \bH(t) \bH^{\dag}(t) \big]$ over time.
Fig. \ref{fig:timeVarying:power} shows the evolution of the users' total transmit power under Algorithm \ref{alg:XLPC} (dashed lines represent the users' maximum transmit power),
while Fig.~\ref{fig:timeVarying:averageRate} shows the achieved/target rate gap $R(t)/R^{\ast}$.
Finally, as in the static channel case, Fig.~\ref{fig:timeVarying:regret} shows the users' average regret $\Reg(\timeHorizon)/\timeHorizon$:
as predicted by Theorem \ref{thm:RegretBound}, the users' regret quickly becomes negative, indicating that their transmit policy is asymptotically optimal in hindsight (for simplicity, we only plot the positive part of the regret and we use a logarithmic dB scale for consistency).%
}
\label{fig:timeVarying}
\end{figure}

Finally, in Fig.~\ref{fig:timeVarying}, we simulate a realistic time-varying environment where the focal transmitters move at different speeds.
For simulation purposes, we used the \ac{EPA}, \ac{EVA}, and \ac{ETU} channel models for pedestrian ($2 \mathrm{km}/\mathrm{h}$), urban vehicular ($30 \mathrm{km}/\mathrm{h}$) and high speed ($130 \mathrm{km}/\mathrm{h}$) users respectively \cite{_user_2014}.
To illustrate the variability of the users' channels, we plot their evolving channel gains $\motr\big[\bH(\dtime) \bH^{\dag}(\dtime) \big]$ in Fig.~\eqref{fig:timeVarying:channel}:
as can be seen, channel variations are quite wide and become more profound for higher user velocities.

In this dynamic setting, the main challenge for the users is to track the optimum signal covariance profile that balances their transmit power against their achieved throughput (i.e. that minimizes their loss) as this optimum profile evolves over time.
To that end, Fig. \ref{fig:timeVarying:power} shows that the users' radiated power under Algorithm \ref{alg:XLPC} increases (to compensate for poor channel conditions) or decreases (when channel conditions are more favorable) in a way consistent with the evolution of the wireless medium (Fig. \ref{fig:timeVarying:channel}).
Dually, in Fig. \ref{fig:timeVarying:averageRate} we plot the time average of the users' achieved/target rate ratio:%
\footnote{Time-averages are considered in order to weed out stochastic fluctuations (due to the users' changing fading environment) that could be potentially misleading.}
as can be seen, users consistently achieve their target throughput, and their achieved/target throughput ratio converges to $1$ over time (in practice, within a few frames for users that do not move at very high speeds).
Furthermore, we see that connections with a softer tolerance for the satisfaction of their \ac{QoS} requirements (e.g. Connection $1$) are very aggressive in reducing transmit power when channel conditions seem to allow it, whereas connections that are less tolerant with respect to their \ac{QoS} requirements (e.g. Connection $2$) are more conservative and transmit at relatively high powers (resulting in higher rates) as a precaution against deep fading events.


Finally, as in the static channel case, Fig. \ref{fig:timeVarying:regret} depicts the users’ average regret over time:
again, despite the pessimistic high-power initialization of Algorithm 1, the users’ regret drops to the no-regret regime in just a few frames (much faster than the $\bigoh(1/\timeHorizon)$ bounds predicted by Theorem \ref{thm:RegretBound}).
The reason for this faster convergence is that the worst-case bounds of Theorem \ref{thm:RegretBound} only become relevant under very adverse (or adversarial) environments, occuring for example when users are being jammed by a third party:
in standard mobility scenarios (such as the one simulated here), the evolution of the wireless medium is relatively tame from a statistical perspective, so users adapt to its variability much faster than in the adversarial regime.

\section{Conclusions}
\label{sec:Conclusions}

In this paper, we examined the trade-off between radiated power and achieved throughput in wireless \ac{MIMO}\textendash\ac{OFDMA} systems that evolve dynamically over time as the result of changing channel conditions and user \ac{QoS} requirements.
To account for the system's complete lack of stationarity (or any other type of average behavior that could allow the use of traditional solution concepts such as Nash/correlated equilibria), we provided a formulation based on online optimization and we derived an adaptive \acl{MXL} algorithm that leads to \emph{no regret} \textendash\ i.e. that is asymptotically optimal in hindsight, irrespective of how the wireless system varies with time.
Importantly, the proposed algorithm requires only local \ac{CSIT} and is robust with respect to measurement errors and imperfections:
in particular, under fairly mild hypotheses for the uncertainty statistics, the proposed algorithm retains its regret minimization properties and converges to a no-regret state.
As a result, thanks to the algorithm's no regret property, the system's users are able to track their optimal transmit power profile ``on the fly'', even under randomly changing channel conditions and high uncertainty.

The proposed algorithmic framework can be readily extended to different precoding schemes (such as MMSE or ZF-type precoders), or to account for other transmission features such as spectral mask constraints, pricing, etc.
Through judicious use of convexification techniques, it can also be applied to non-convex energy-efficiency objectives, such as the users' achieved throughput per Watt of radiated power;
we intend to explore these directions in future work.


\appendix[Technical Proofs]
\label{app:proofs}

Our goal in this appendix is to prove the regret guarantees of \eqref{eq:MXL} under both perfect and imperfect \ac{CSI} (Theorems \ref{thm:RegretBound} and \ref{thm:ExpectedRegretBound} respectively).
Drawing on the approach of \cite{Sor09,KM14}, we will first establish the no-regret properties of Algorithm \ref{alg:XLPC} in a continuous-time, ``mean-field'' setting, and we will then show that these properties descend to discrete time at the cost of an extra term in the algorithm's regret guarantees.
The algorithm's robustness properties with respect to measurement noise and errors will then follow by using the theory of concentration inequalities.

For notational clarity and convenience, we will be suppressing the dependence on time whenever possible, and we will write e.g. $\dot\bQ$ instead of $\frac{d}{d\ctime} \power{\ctime}$ when there is no ambiguity.

\subsection{No regret in continuous time}
\label{sec:ProofOfRegretBoundContinuousTime}

We begin by considering the following continuous-time analogue of the basic \acl{PC} algorithm \eqref{eq:MXL}:
\begin{equation}
\label{eq:MXL-cont}
\tag{MXL-c}
\begin{aligned}
\dot\bY
	&= - \bV,
	\\
\power{}
	&= \pmax \frac{\exp(\temp(\ctime) \bY)}{1 + \motr\big[ \exp(\temp(\ctime) \bY) \big]},
\end{aligned}
\end{equation}
where $\temp(\ctime) > 0$ is a smooth, nonincreasing learning parameter and the gradient matrix $\bV$ is defined as in \eqref{eq:LossDerivative}.
The following proposition shows that \eqref{eq:MXL-cont} leads to no regret in continuous time:

\begin{proposition}
\label{prop:NoRegretContinuousTime}
The learning scheme \eqref{eq:MXL-cont} guarantees the continuous-time regret bound:
\begin{equation}
\label{eq:RegretBoundContinuousTime}
\max_{\fxPower\in\state}
	\int_{0}^{\timeHorizon} \big[ \loss{\power{\ctime}}{\ctime} - \loss{\fxPower}{\ctime} \big] \dd\ctime
	\leq \frac{\pmax\cdot\log(1 + \nOfSubcarriers\nOfTx)}{\temp(\timeHorizon)},
\end{equation}
for any measurable stream of effective channel matrices $\effChannel{\ctime}$, $\ctime\geq0$.
In particular, if $\temp(\ctime)$ satisfies the decay rate condition $\lim_{t\to\infty} \ctime\cdot\temp(\ctime) = \infty$, the learning scheme \eqref{eq:MXL-cont} leads to no regret.
\end{proposition}

\begin{IEEEproof}
We first note that the loss function $\loss{\power{}}{\ctime}$ is convex with respect to $\power{}$ (to see this, simply recall that the Shannon rate function $\rate{\power{}}{\ctime}$ is concave in $\power{}$ \cite{boyd_convex_2004} while $\phi$ is assumed concave and nondecreasing).
With this in mind, we obtain:
\begin{equation}
\label{eq:monotonicity}
\loss{\power{\ctime}}{\ctime} - \loss{\fxPower}{\ctime}
	\leq \motr\left[ \left( \power{\ctime} - \fxPower \right) \cdot \gradient{\ctime} \right],
\end{equation}
where $\gradient{\ctime} = \nabla_{\power{\ctime}}\loss{\power{\ctime}}{\ctime}$ denotes the gradient of $\loss{\cdot}{\ctime}$ evaluated at $\power{\ctime}$.
Accordingly, to establish the no-regret bound \eqref{eq:RegretBoundContinuousTime} for \eqref{eq:MXL}, it suffices to show that
\begin{equation}
\label{eq:LossBound:0}
\int_{0}^{\timeHorizon} \motr\left[ \left( \power{\ctime} - \fxPower \right) \cdot \gradient{\ctime} \right] \dd\ctime
	\leq \frac{\pmax\cdot\log(1 + \nOfSubcarriers\nOfTx)}{\temp(\timeHorizon)}
\end{equation}
for all $\fxPower\in\state$.

To that end, \eqref{eq:MXL} readily yields:
\begin{flalign}
\label{eq:LossBound:1}
\int_{0}^{\timeHorizon} \motr\left[ \left( \power{\ctime} - \fxPower \right) \cdot \gradient{\ctime} \right] \,d\ctime
	&= \int_{0}^{\timeHorizon} \motr\big[ ( \fxPower - \power{\ctime}) \cdot \dot\bY(\ctime) \big] \,d\ctime
	\notag\\
	&= \motr\big[ \bY(\timeHorizon) \cdot \fxPower \big]
	- \int_{0}^{\timeHorizon} \motr\big[ \power{\ctime} \dot\bY(\ctime) \big] \dd\ctime,
\end{flalign}
where we have used the fact that $\bY(0) = 0$.
To continue, note that the exponentiation step of \eqref{eq:MXL} can be written more simply as:
\begin{equation}
\label{eq:mirror}
\power{}
	= \pmax \cdot \nabla_{\bU} \log\left[ 1 + \motr \exp(\bU)  \right],
\end{equation}
where we have set $\bU = \temp\bY$.%
\footnote{This is actually one of the main reasons behind the exponentiation step of \eqref{eq:MXL}.}
As a result, with $\dot\bU = \dot\temp\bY + \temp\dot\bY$, the integrand of the second term of \eqref{eq:LossBound:1} becomes:
\begin{equation}
\label{eq:LossBound:2}
\motr\big[ \power{} \dot \bY \big]
	= \frac{1}{\temp} \motr\big[ \power{} \dot \bU \big]
	- \frac{\dot\temp}{\temp^{2}} \motr\big[ \power{} \bU \big]
	= \frac{\pmax}{\temp} \frac{d}{d\ctime} \log\left[ 1 + \motr \exp(\bU) \right]
	- \frac{\dot\temp}{\temp^{2}} \motr\big[ \power{} \bU \big].
\end{equation}
Hence, after integrating \eqref{eq:LossBound:2} by parts (and recalling that $\bU(0) = 0$), we get:
\begin{flalign}
\label{eq:LossBound:3}
\int_{0}^{\timeHorizon} \motr\big[ \power{\ctime} \dot\bY(\ctime) \big] \dd\ctime
	&= \left. \frac{\pmax}{\temp(\ctime)} \log\left[ 1 + \motr \exp(\bU(\ctime)) \right] \right\vert_{0}^{\timeHorizon}
	\notag\\
	&+ \pmax\int_{0}^{\timeHorizon} \frac{\dot\temp(\ctime)}{\temp(\ctime)^{2}} \log\left[ 1 + \motr\big[\exp(\bU(\ctime))\big] \right] d\ctime
	- \int_{0}^{\timeHorizon} \frac{\dot\temp(\ctime)}{\temp(\ctime)^{2}} \motr\big[\power{\ctime} \bU(\ctime) \big] \dd\ctime
	\notag\\
	&= \frac{\pmax \log\left[1 + \motr \exp(\bU(\timeHorizon))\right]}{\temp(\timeHorizon)}
	- \frac{\pmax \log(1 + \nOfSubcarriers\nOfTx)}{\temp(0)}
	\notag\\
	&+ \int_{0}^{\timeHorizon} \frac{\dot\temp(\ctime)}{\temp(\ctime)^{2}} \Big[
	\pmax \log\left[ 1 + \motr \exp(\bU(\ctime)) \right] - \motr[\power{\ctime} \bU(\ctime)]
	\Big] \dd\ctime,
\end{flalign}
where we have used the fact that $\bU(0) = 0$ (implying in turn that $\motr\exp(\bU(0)) = \nOfSubcarriers \nOfTx$).
Thus, combining all of the above, we obtain:
\begin{flalign}
\label{eq:LossBound:4}
\int_{0}^{\timeHorizon} \motr\left[ \left( \power{\ctime} - \fxPower \right) \cdot \gradient{\ctime} \right] d\ctime
	&= \frac{\pmax \log(1 + \nOfSubcarriers\nOfTx)}{\temp(0)}
	+ \frac{\motr\left[ \fxPower\bU(\timeHorizon) \right] - \pmax \log\left[1 + \motr\exp(\bU(\timeHorizon)) \right]}{\temp(\timeHorizon)}
	\notag\\
	&+ \int_{0}^{\timeHorizon} \frac{\dot\temp(\ctime)}{\temp(\ctime)^{2}} \Big[
	\motr[\power{\ctime} \bU(\ctime)] - \pmax\log\left[ 1 + \motr \exp(\bU(\ctime)) \right]
	\Big] \dd\ctime,
\end{flalign}

To proceed, we will require the inequality:
\begin{equation}
\label{eq:Fenchel}
\motr[\bA\bX] - \log\big[ 1 + \motr\exp(\bX) \big]
	\leq \motr[\bA \log \bA] + (1-\motr\bA) \log\left(1 - \motr\bA \right)
\end{equation}
valid for all Hermitian $\bA,\bX$, with $\bA\mgeq 0$, $\motr\bA \leq 1$, and with equality holding if and only if
\begin{equation}
\label{eq:choice}
\bA
	= \frac{\exp(\bX)}{1 + \motr\exp(\bX)}.
\end{equation}
To establish \eqref{eq:Fenchel}, it clearly suffices to show that the supremum of its LHS for fixed $\bA$ is precisely the RHS of \eqref{eq:Fenchel}.
Accordingly, let
\begin{equation}
F(\bX)
	= \log\big[ 1 + \motr\exp(\bX) \big] - \motr[\bA\bX],
\end{equation}
so the maximizers of the LHS of \eqref{eq:Fenchel} are given by the first-order stationarity condition $\nabla_{\bX} F(\bX) = 0$ (simply note that $F(\bX)$ is strictly concave in $\bX$).
By differentiating, we then obtain:
\begin{equation}
\label{eq:dF}
\nabla_{\bX} F(\bX)
	= \frac{\exp(\bX)}{1 + \motr\exp(\bX)} - \bA.
\end{equation}
Thus, if $\bA\mg0$ and $\tr\bA < 1$, the equation $\nabla_{\bX} F(\bX) = 0$ always admits a (necessarily unique) solution given by:
\begin{equation}
\eq
	= \log\bA + \log(1+\chi) \eye,
\end{equation}
with $\chi = \motr\exp(\eq)$.
Moreover, setting $a=\motr\bA$ and tracing \eqref{eq:dF} readily yields $\chi = a/(1-a)$, so, after some easy algebra, the maximum value of $F$ will be:
\begin{flalign}
F_{\max}
	&= F(\eq)
	= \motr\big[ \bA \log\bA \big] + (1 - a) \log(1 - a).
\end{flalign}
The above establishes \eqref{eq:Fenchel} for the case $\bA\mg0$ and $\tr\bA < 1$;
the boundary cases $\det\bA=0$ and/or $\tr\bA=1$ then follow by continuity.

Thus, returning to \eqref{eq:LossBound:4}, an immediate application of \eqref{eq:Fenchel} gives:
\begin{subequations}
\label{eq:Fenchel:2}
\begin{flalign}
\label{eq:Fenchel:2a}
\motr\left[ \fxPower\bU(\timeHorizon) \right] - \pmax \log\left[1 + \motr\exp(\bU(\timeHorizon)) \right]
	&\leq 0,
	\\
\label{eq:Fenchel:2b}
\motr[\power{\ctime} \bU(\ctime)] - \pmax\log\left[ 1 + \motr \exp(\bU(\ctime)) \right]
	&= \pmax\cdot\left[
	\motr[\bA(\ctime) \log\bA(\ctime)] + (1 - a(\ctime)) \log(1 - a(\ctime))
	\right],
\end{flalign}
\end{subequations}
where we have set $\bA(\ctime) = \power{\ctime}/\pmax$ and $a(\ctime) = \motr\bA(\ctime)$.
As for \eqref{eq:Fenchel:2b}, its RHS can be expressed more concisely as the (negative) von Neumann quantum entropy of the augmented matrix $\bA_{0}(\ctime) = \diag{a(\ctime),\bA(\ctime)}$, i.e.
\begin{equation}
\label{eq:maxentropy}
\motr[\bA(\ctime) \log\bA(\ctime)] + (1 - a(\ctime)) \log(1 - a(\ctime))
	= \motr[\bA_{0}(\ctime) \log\bA_{0}(\ctime)]
	\geq -\log(1 + \nOfSubcarriers\nOfTx),
\end{equation}
where the last inequality simply corresponds to the maximum value of the von Neumann entropy (recall also that $\dim(\bA_{0}) = 1 + \nOfSubcarriers\nOfTx$) \cite{carlen_trace_2009}.
Thus, substituting \eqref{eq:Fenchel:2} and \eqref{eq:maxentropy} back into \eqref{eq:LossBound:4}, we finally obtain:

\begin{flalign}
\label{eq:LossBound:5}
\int_{0}^{\timeHorizon} \motr\left[ \left( \power{\ctime} - \fxPower \right) \cdot \gradient{\ctime} \right] d\ctime
	&\leq \pmax \log(1 + \nOfSubcarriers\nOfTx)
	\left[
	\frac{1}{\temp(0)} - \int_{0}^{\timeHorizon} \frac{\dot\temp(\ctime)}{\temp(\ctime)^{2}} \dd\ctime
	\right]
	= \frac{\pmax \log(1 + \nOfSubcarriers\nOfTx)}{\temp(\timeHorizon)},
\end{flalign}
where we have used the fact that $\dot\temp \leq 0$ (recall that $\temp$ has been assumed nonincreasing).
The regret bound \eqref{eq:RegretBoundContinuousTime} then follows by maximizing \eqref{eq:monotonicity} over all $\fxPower\in\state$.
\end{IEEEproof}

\subsection{No regret in discrete time: the case of perfect \ac{CSI}}
\label{sec:ProofOfRegretBoundWithPerfectCSI}

We now return to the discrete-time process \eqref{eq:MXL}, written here in the more general form:
\begin{equation}
\begin{aligned}
\cumGradient{\dtime}
	&= -\insum_{\dtimeAux=1}^{\dtime}\gradient{\dtimeAux}
	\\
\power{\dtime+1}
	&= \pmax \frac{\exp\left(\etaSeq{\dtime}\cumGradient{\dtime}\right)}{1 + \motr\left[ \exp\left(\etaSeq{\dtime}\cumGradient{\dtime}\right)\right]}
  \end{aligned}
  \label{eq:dXL}
\end{equation}
with $\temp(\dtime) = \temp\dtime^{-1/2}$ for some positive parameter $\temp>0$.
To establish the regret bound \eqref{eq:RegretBound} of Theorem \ref{thm:RegretBound}, we will define an interpolated continuous-time process, use Proposition \ref{prop:NoRegretContinuousTime} to estimate the incurred regret in continuous time, and use a discrete-continuous comparison argument in order to bound the regret in discrete time.

\begin{IEEEproof}[Proof of Theorem \ref{thm:RegretBound}]
We begin by constructing a continuous-time interpolation of \eqref{eq:MXL} and comparing it to its discrete-time analogue.
To that end, consider the continuous-time processes $\gradientC{\ctime} = \gradient{\ceil{\ctime}}$ and $\etaSeqC{\ctime} = \etaSeq{\floor{\ctime}}$ for all $\ctime\geq0$, with $\etaSeqC{0} = \etaSeq{0} = \temp$ by convention.%
\footnote{Note that $\gradientC{\ctime} = \gradient{\dtime}$ and $\etaSeqC{\ctime} = \etaSeq{\dtime-1}$ for all $\ctime\in\left(\dtime-1, \dtime\right)$, i.e. $\gradientC{\ctime}$ precedes its discrete-time analogue, while $\etaSeqC{\ctime}$ lags behind it;
this one-step offset will be key in the rest of our proof.}
In this context, the continuous-time learning scheme \eqref{eq:MXL-cont} yields the processes:
\begin{equation}
\begin{aligned}
\cumGradientC{\ctime}
	&= -\int_{0}^{\ctime}\gradientC{\ctimeAux}d\ctimeAux
	\\
\powerC{\ctime}
	&= \pmax \frac{\exp\inPar{\etaSeqC{\ctime}\cumGradientC{\ctime}}}{1+\tr{\exp\inPar{\etaSeqC{\ctime}\cumGradientC{\ctime}}}}
\end{aligned}
\label{eq:dXL:contEquiv}
\end{equation}
whence we easily obtain:
\ifCLASSOPTIONdraftcls
\begin{equation}
\begin{aligned}
\cumGradientC{\dtime}
    	&=  -\int_{0}^{\dtime}\gradientC{\ctimeAux}d\ctimeAux
	= -\insum_{\dtimeAux=1}^{\dtime}{\int_{\dtimeAux-1}^{\dtimeAux}}\gradient{\dtimeAux}d\ctimeAux
	= -\insum_{\dtimeAux=1}^{\dtime}\gradient{\dtimeAux} =\cumGradient{\dtime},
	\\
\powerC{\dtime}
	&= \power{\dtime+1},
\end{aligned}
\end{equation}
\else
\begin{equation}
\begin{aligned}
\cumGradientC{\dtime}
	&=  -\int_{0}^{\dtime}\gradientC{\ctimeAux}d\ctimeAux
	\\
	&= -\insum_{\dtimeAux=1}^{\dtime}{\int_{\dtimeAux-1}^{\dtimeAux}}\gradient{\dtimeAux}d\ctimeAux
	= -\insum_{\dtimeAux=1}^{\dtime}\gradient{\dtimeAux}
	\\
 	&=\cumGradient{\dtime}
	\\
\powerC{\dtime}
	&= \power{\dtime+1}
\end{aligned}
\end{equation}
\fi
%
Consequently, for all $\dtime \geq 1$ and for all $\ctime\in\inPar{\dtime-1,\dtime}$, H\"older's inequality yields:
\ifCLASSOPTIONdraftcls
\begin{flalign}
\abs{ \motr\left[ \inPar{\powerC{\ctime}-\power{\dtime}} \cdot \gradientC{\ctime} \right] }
	&= \abs{ \motr\left[ \inPar{\powerC{\ctime}-\powerC{\dtime-1}} \cdot \gradientC{\ctime} \right] }
	\notag\\
	&\leq \norm{\gradientC{\ctime}} \cdot \abs{ \motr\left[ \powerC{\ctime}-\powerC{\dtime-1} \right] }
	\leq \gradientB \cdot \abs{\motr\left[ \powerC{\ctime}-\powerC{\dtime-1} \right] }
\label{eq:LossBound:Discrete:1}
\end{flalign}
\else
\begin{flalign}
	&\abs{\tr{\inPar{\powerC{\ctime}-\power{\dtime}} \gradientC{\ctime}}}
	=\abs{\tr{\inPar{\powerC{\ctime}-\powerC{\dtime-1}} \gradientC{\ctime}}}
	\notag\\
	&\quad
	\leq \norm{\gradientC{\ctime}}\tr{\powerC{\ctime}-\powerC{\dtime-1}}
	\leq \gradientB\tr{\powerC{\ctime}-\powerC{\dtime-1}}
\label{eq:LossBound:Discrete:1}
\end{flalign}
\fi
Using the analysis of \cite{KSST12}, it can be shown that the map $\bU\mapsto \pmax \exp(\bU) \big/ [1 + \motr \exp(\bU)]$ is $(\pmax/2)$\textendash Lip\-schitz with respect to the spectral and nuclear norms (for the map's domain and codomain respectively).
We may thus write:
\begin{equation}
\motr\left[ \powerC{\ctime}-\powerC{\dtime-1} \right]
	\leq \tfrac{1}{2} \pmax\,
	\norm{\etaSeqC{\ctime}\cumGradientC{\ctime} - \etaSeqC{\dtime-1}\cumGradientC{\dtime-1}}
	= \tfrac{1}{2} \etaSeq{\dtime-1} \pmax \,
	\norm{\cumGradientC{\ctime} - \cumGradientC{\dtime-1}}.
	\!
\label{eq:LossBound:Discrete:2}
\end{equation}
Furthermore, by definition, we also have:
\ifCLASSOPTIONdraftcls
\begin{flalign}
\norm{\cumGradientC{\ctime} - \cumGradientC{\dtime-1}}
	&= \norm{-\int_{0}^{\ctime}\gradientC{\ctimeAux} \dd\ctimeAux + \int_{0}^{\dtime-1}\gradientC{\ctimeAux} \dd\ctimeAux}
	\leq \int_{\dtime-1}^{\ctime} \norm{\gradientC{\ctimeAux}} d\ctimeAux
	\leq \gradientB \cdot \inPar{\ctime-\dtime+1}
\label{eq:LossBound:Discrete:3}
\end{flalign}
\else
\begin{equation}
  \begin{aligned}
   & \norm{\cumGradientC{\ctime} - \cumGradientC{\dtime-1}} = \norm{-\int_{0}^{\ctime}\gradientC{\ctimeAux}d\ctimeAux + \int_{0}^{\dtime-1}\gradientC{\ctimeAux}d\ctimeAux} \\
   & \quad = \norm{-\int_{\dtime-1}^{\ctime}\gradientC{\ctimeAux}d\ctimeAux} = \norm{-\int_{\dtime-1}^{\ctime}\gradient{\dtime}d\ctimeAux} \\
   & \quad = \norm{\gradient{\dtime}}\inPar{\ctime-\dtime+1} \leq \gradientB\inPar{\ctime-\dtime+1}
  \end{aligned}
  \label{eq:LossBound:Discrete:3}
\end{equation}
\fi
and hence, by combining \eqref{eq:LossBound:Discrete:1}, \eqref{eq:LossBound:Discrete:2} and \eqref{eq:LossBound:Discrete:3}, we get:
\begin{equation}
\abs{ \motr\left[ \inPar{\powerC{\ctime}-\power{\dtime}} \gradientC{\ctime} \right] }
	\leq \frac{1}{2}\pmax\norm{\gradient{\dtime}}^{2} \,\etaSeq{\dtime-1} \cdot \inPar{\ctime-\dtime+1}.
\label{eq:LossBound:Discrete:5}
\end{equation}

Accordingly, with this discrete/continuous comparison result at hand, we get:
\ifCLASSOPTIONdraftcls
\begin{flalign}
\bigg\vert
	\int_{0}^{\timeHorizon}{\tr{\powerC{\ctime}\gradientC{\ctime}}} \dd\ctime
	&- \insum_{\dtime=1}^{\timeHorizon} \motr\left[\power{\dtime}\gradient{\dtime} \right]
	\bigg\vert
	= \abs{\insum_{\dtime=1}^{T}\int_{\dtime-1}^{\dtime} \motr\left[ \powerC{\ctime}\gradientC{\ctime} \right] d\ctime - \motr\left[\power{\dtime}\gradient{\dtime} \right] }
	\notag\\ 
	&= \abs{
	\insum_{\dtime=1}^{T} \int_{\dtime-1}^{\dtime} \big[
		\motr\left( \powerC{\ctime}\gradientC{\ctime} \right)
		-\motr\left( \power{\dtime}\gradient{\dtime} \right)
	\big] \dd\ctime }
	\notag\\
	&\leq \insum_{\dtime=1}^{T}\int_{\dtime-1}^{\dtime}{\abs{\tr{\powerC{\ctime}\gradientC{\ctime}}-\tr{\power{\dtime}\gradientC{\ctime}}}d\ctime}
	\notag\\
	&\leq \insum_{\dtime=1}^{T}\int_{\dtime-1}^{\dtime}{\frac{\pmax}{2} \norm{\gradient{\dtime}}^{2} \etaSeq{\dtime-1}\inPar{\ctime-\dtime+1}}d\ctime
	\leq \frac{\pmax\gradientB^2}{4}\insum_{\dtime=1}^{T}\etaSeq{\dtime-1},
\label{eq:LossBound:Discrete:6}
\end{flalign}
\else
\begin{equation}
  \begin{aligned}
    & \abs{\int_{0}^{\timeHorizon}{\tr{\powerC{\ctime}\gradientC{\ctime}}}d\ctime - \insum_{\dtime=1}^{T}{\tr{\power{\dtime}\gradient{\dtime}}}} \\
    & \quad = \abs{\insum_{\dtime=1}^{T}\int_{\dtime-1}^{\dtime}{\tr{\powerC{\ctime}\gradientC{\ctime}}}d\ctime - \insum_{\dtime=1}^{T}{\tr{\power{\dtime}\gradient{\dtime}}}} \\ 
    & \quad = \abs{\insum_{\dtime=1}^{T}\int_{\dtime-1}^{\dtime}{\inPar{\tr{\powerC{\ctime}\gradientC{\ctime}}-\tr{\power{\dtime}\gradient{\dtime}}}d\ctime} } \\
    & \quad \leq \insum_{\dtime=1}^{T}\int_{\dtime-1}^{\dtime}{\abs{\tr{\powerC{\ctime}\gradientC{\ctime}}-\tr{\power{\dtime}\gradientC{\ctime}}}d\ctime} \\
    & \quad \leq \insum_{\dtime=1}^{T}\int_{\dtime-1}^{\dtime}{\frac{\pmax}{2}\gradientB^2\etaSeq{\dtime-1}\inPar{\ctime-\dtime+1}}d\ctime \\
    & \quad = \insum_{\dtime=1}^{T}\frac{\pmax}{2}\gradientB^2\etaSeq{\dtime-1}\frac{1}{2} = \frac{\pmax\gradientB^2}{4}\insum_{\dtime=1}^{T}\etaSeq{\dtime-1}
  \end{aligned}
\end{equation}
\fi
for all $\timeHorizon \geq 1$.
Thus, using Proposition \ref{prop:NoRegretContinuousTime} and the convexity condition \eqref{eq:monotonicity}, we obtain:
\ifCLASSOPTIONdraftcls
\begin{flalign}
\insum_{\dtime=1}^{\timeHorizon}\inBra{\loss{\power{\dtime}}{\dtime} - \loss{\fxPower}{\dtime}}
	&\leq \insum_{\dtime=1}^{\timeHorizon} \motr\left[ \inPar{\power{\dtime}-\fxPower}\gradient{\dtime} \right]
	\notag\\
	&\leq \int_{0}^{\timeHorizon} \motr\left[ \powerC{\ctime}\gradientC{\ctime} \right] d\ctime
	+ \frac{\pmax\gradientB^{2}}{4}\insum_{\dtime=1}^{\timeHorizon} \etaSeq{\dtime-1}
	- \int_{0}^{\timeHorizon} \motr\left[ \fxPower\gradientC{\ctime} \right] d\ctime
	\notag\\
	&= \int_{0}^{\timeHorizon} \motr\left[ \inPar{\powerC{\ctime}-\fxPower}\gradientC{\ctime} \right] d\ctime
	+\frac{\pmax\gradientB^2}{4}\insum_{\dtime=1}^{T}\etaSeq{\dtime-1}
	\notag\\
	&\leq \frac{\pmax\log\inPar{1+\nOfSubcarriers\nOfTx}}{\etaSeq{\timeHorizon}}
	+ \frac{\pmax\gradientB^2}{4}\insum_{\dtime=1}^{T}\etaSeq{\dtime-1},
\label{eq:LossBound:Discrete:7}
\end{flalign}
\else
\begin{equation}
  \begin{aligned}
    \regret{\fxPower}{\timeHorizon} & = \insum_{\dtime=1}^{\timeHorizon}\inPar{\loss{\power{\dtime}}{\dtime} - \loss{\fxPower}{\dtime}} \\
    & \leq \insum_{\dtime=1}^{\timeHorizon}{\tr{\inPar{\power{\dtime}-\fxPower}\gradient{\dtime}}} \\ & 
    =  \insum_{\dtime=1}^{\timeHorizon}{\tr{\power{\dtime}\gradient{\dtime}}}-\insum_{\dtime=1}^{\timeHorizon}\tr{\fxPower\gradient{\dtime}} \\
    & \leq \int_{0}^{\timeHorizon}{\tr{\powerC{\ctime}\gradientC{\ctime}}}d\ctime+\frac{\pmax\gradientB^2}{4}\insum_{\dtime=1}^{T}\etaSeq{\dtime-1} \\
    & \quad - \int_{0}^{\timeHorizon}{\tr{\fxPower\gradientC{\ctime}}}d\ctime\\
    & = \int_{0}^{\timeHorizon}{\tr{\inPar{\powerC{\ctime}-\fxPower}\gradientC{\ctime}}}d\ctime
    +\frac{\pmax\gradientB^2}{4}\insum_{\dtime=1}^{T}\etaSeq{\dtime-1} \\
    & \leq \frac{\pmax\log\inPar{1+\nOfSubcarriers\nOfTx}}{\etaSeq{\timeHorizon}} + \frac{\pmax\gradientB^2}{4}\insum_{\dtime=1}^{T}\etaSeq{\dtime-1}
  \end{aligned}
\end{equation}
\fi
where we used \eqref{eq:LossBound:Discrete:6} and the fact that $\insum_{\dtime=1}^{\timeHorizon} \motr\left[ \fxPower\gradient{\dtime} \right] = \int_{0}^{\timeHorizon} \motr\left[ \fxPower\gradientC{\ctime} \right] d\ctime$ in the second line.
Thus, substituting $\etaSeq{\dtime} = \min\{\etaSqrt{\dtime},\eta\}$, the last term of \eqref{eq:LossBound:Discrete:7} becomes
\ifCLASSOPTIONdraftcls
\begin{equation}
\insum_{\dtime=1}^{\timeHorizon} \etaSeq{\dtime-1}
	= \eta + \insum_{\dtime=1}^{\timeHorizon-1}\etaSqrt{\dtime}
	\leq \eta + \int_{0}^{\timeHorizon-1}\etaSqrt{\ctime}d\ctime \leq \eta\inPar{1+2\sqrt{T}},
\end{equation}
\else
\begin{equation}
  \begin{aligned}
    & \insum_{\dtime=1}^{T}\etaSeq{\dtime-1} = \etaSeq{0} + \insum_{\dtime=2}^{T}\etaSeq{\dtime-1} = \etaSeq{1} + \insum_{\dtime=1}^{T-1}\etaSeq{\dtime} \\
    & \quad = \eta + \insum_{\dtime=1}^{T-1}\etaSqrt{\dtime} \leq \eta + \int_{0}^{\timeHorizon-1}\etaSqrt{\ctime}d\ctime \leq \eta\inPar{1+2\sqrt{T}}\\
  \end{aligned}
\end{equation}
\fi
and our proof is completed by substituting in \eqref{eq:LossBound:Discrete:7} and maximizing over all $\fxPower\in\state$.
\end{IEEEproof}

\subsection{No regret in discrete time: the case of imperfect \ac{CSI}}
\label{sec:ProofOfRegretBoundWithImperfectCSI}

To prove Theorem \ref{thm:ExpectedRegretBound}, we will use Eq.~\eqref{eq:LossBound:Discrete:6} to bound the user's ``virtual'' regret with respect to the sequence of noisy gradient estimates $\gradientN{\dtime}$, and we will then employ the Borel\textendash Cantelli lemma to show that the user's actual regret lies within a vanishing window of his ``virtual'' regret.

\begin{IEEEproof}[Proof of Theorem \ref{thm:ExpectedRegretBound}]
As usual, the user's regret is bounded by:
\begin{equation}
\label{eq:LossBoundStoch:1}
\Reg(\timeHorizon)
	= \max_{\fxPower\in\state} \insum_{\dtime=1}^{\timeHorizon}
	\left[ \loss{\power{\dtime}}{\dtime} - \loss{\fxPower}{\dtime} \right]
	\leq \max_{\fxPower\in\state} \insum_{\dtime=1}^{\timeHorizon}
	\motr\left[ (\power{\dtime} - \fxPower) \cdot \gradient{\dtime} \right],
\end{equation}
so, for the first part of the theorem, it suffices to show that $\insum_{\dtime=1}^{\timeHorizon} \motr\left[ (\power{\dtime} - \fxPower) \cdot \gradient{\dtime} \right] = o(\timeHorizon)$ for all $\fxPower\in\state$.
To that end, given that $\gradient{\dtime} = \gradientN{\dtime} - \noiseGradient{\dtime}$, we have:
\begin{equation}
\label{eq:LossBoundStoch:2}
\motr\left[ (\power{\dtime} - \fxPower) \cdot \gradient{\dtime} \right]
	= \motr\big[ (\power{\dtime} - \fxPower) \cdot \gradientN{\dtime} \big]
	- \motr\left[ (\power{\dtime} - \fxPower) \cdot \noiseGradient{\dtime} \right],
\end{equation}
where $\power{\dtime}$ is defined via the stochastic recursion:
\begin{equation}
\label{eq:MXL-noisy}
\begin{aligned}
\cumGradient{\dtime}
	&= \cumGradient{\dtime-1} - \gradientN{\dtime},
	\\
\power{\dtime+1}
	&= \pmax \frac{\exp( \etaSqrt{\dtime}\cumGradient{\dtime})}{1+\motr\big[ \exp(\etaSqrt{\dtime} \cumGradient{\dtime} \big]}.
\end{aligned}
\end{equation}
Going back to the proof of Thm.~\ref{thm:RegretBound}, we may then use the last inequality of \eqref{eq:LossBound:Discrete:6} to rewrite \eqref{eq:LossBound:Discrete:7} as:
\begin{equation}
\label{eq:LossBoundStoch:3}
\insum_{\dtime=1}^{\timeHorizon}
	\motr\big[ (\power{\dtime} - \fxPower) \cdot \gradientN{\dtime} \big]
	\leq \frac{\pmax\log\inPar{1+\nOfSubcarriers\nOfTx}}{\etaSeq{\timeHorizon}}
	+ \frac{\pmax}{4} \insum_{\dtime=1}^{\timeHorizon} \temp(\dtime-1) \norm{\gradientN{\dtime}}^{2}.
\end{equation}
The last term of \eqref{eq:LossBoundStoch:3} can then be bounded as:
\begin{flalign}
\insum_{\dtime=1}^{\timeHorizon} \temp(\dtime-1) \norm{\gradientN{\dtime}}^{2}
	&\leq \insum_{\dtime=1}^{\timeHorizon} \temp(\dtime-1)
	\left[
	\norm{\gradient{\dtime}}^{2}
	+ 2 \norm{\gradient{\dtime}}\cdot\norm{\noiseGradient{\dtime}} + \norm{\noiseGradient{\dtime}}^{2}
	\right]
	\notag\\
	&= \gradientB^{2} \insum_{\dtime=1}^{\timeHorizon} \temp(\dtime-1)
	+ \bigoh\left(\insum_{\dtime=1}^{\timeHorizon} \temp(\dtime-1) \norm{\noiseGradient{\dtime}}^{2}\right),
\end{flalign}
where we have used the triangle inequality in the first line.
We now claim that
\begin{equation}
\frac{1}{\timeHorizon} \insum_{\dtime=1}^{\timeHorizon} \temp(\dtime-1) \norm{\noiseGradient{\dtime}}^{2}
	\to 0
	\quad
	\text{as $\timeHorizon\to\infty$ (a.s.).}
\end{equation}
Indeed, if we let $z(\dtime) = \norm{\noiseGradient{\dtime}}$, Hypothesis \eqref{eq:tailbound} implies that $\prob\left(z(\dtime) \geq \dtime^{1/4-\eps} \right) = \bigoh(1/\dtime^{\beta})$ for some $\beta>1$ and for all small enough $\eps>0$.
We thus obtain:
\begin{equation}
\insum_{\dtime=1}^{\infty} \prob\left(z(\dtime) \geq \dtime^{1/4-\eps} \right)
	= \bigoh\left( \insum_{\dtime=1}^{\infty} \dtime^{-\beta} \right)
	= \bigoh(1)
	< \infty,
\end{equation}
and hence, by the Borel\textendash Cantelli lemma, we conclude that
\begin{equation}
\prob\left(
	\text{$z(\dtime) \geq \dtime^{1/4-\eps}$ for infinitely many $\dtime$}
	\right)
	= 0.
\end{equation}
In turn, this implies that $z(\dtime)^{2} = \bigoh\big(\dtime^{1/2-2\eps}\big)$ almost surely, so, with $\temp(\dtime) = \dtime^{-1/2}$, we get:
\begin{equation}
\label{eq:LossBoundStoch:4}
\insum_{\dtime=1}^{\timeHorizon} \temp(\dtime-1) \norm{\noiseGradient{\dtime}}^{2}
	= \bigoh\left( \insum_{\dtime=1}^{\timeHorizon} \dtime^{-1/2} \dtime^{1/2 - 2\eps} \right)
	= \bigoh\left( \insum_{\dtime=1}^{\timeHorizon} 1/\dtime^{2\eps} \right)
	= o(\timeHorizon)
	\quad
	\text{(a.s.)}.
\end{equation}

For the second term of \eqref{eq:LossBoundStoch:2}, let $\xi(\dtime) = \motr\left[ (\power{\dtime} - \fxPower) \cdot \noiseGradient{\dtime} \right]$.
Then, given that $\power{\dtime}$ is fully determined by $\power{\dtime-1}$ and $\noiseGradient{\dtime-1}$, it follows that $\ex\left[\xi(\dtime) \,\vert\, \power{\dtime-1}\right] = 0$, i.e. $\xi(\dtime)$ is a martingale difference sequence;
as a result, we get $\lim_{\timeHorizon\to\infty} \timeHorizon^{-1} \insum_{\dtime=1}^{\timeHorizon} \xi(\dtime) = 0$ by the strong law of large numbers for martingale differences \textendash\ see e.g. Theorem 2.18 in \cite{HH80}.
Combining this with \eqref{eq:LossBoundStoch:4}, we then get
\begin{equation}
\insum_{\dtime=1}^{\timeHorizon} \motr\left[ (\power{\dtime} - \fxPower) \cdot \gradient{\dtime} \right]
	= o(\timeHorizon)
	\quad
	\text{(a.s.),}
\end{equation}
i.e. \eqref{eq:MXL-noisy} leads to no regret, as claimed.
The mean bound \eqref{eq:ExpectedRegretBound} is then obtained by taking expectations on both sides of \eqref{eq:LossBoundStoch:3} and recalling that $\ex[\gradientN{\dtime} \vert \power{n-1}] = \gradient{\dtime}$.
\end{IEEEproof}


\footnotesize
\balance
\bibliographystyle{IEEEtran}
\bibliography{IEEEabrv,Bibliography-PM,MyLibrary}

\end{document}